\documentclass[reprint, superscriptaddress, amsmath, amssymb, aps, prb]{revtex4-2}

\usepackage{graphicx} 
\usepackage{dcolumn}  
\usepackage{bm}       
\usepackage{hyperref} 
\usepackage{float}
\usepackage{subfigure}
\usepackage{amsmath}
\usepackage{comment}

\begin{document}

\title{Impact of Random Bond Disorder on Quantum Skyrmions in a spin-half Quantum Heisenberg Model}

\author{Amit Kumar}
\affiliation{Theory Division, Saha Institute of Nuclear Physics,
A CI of Homi Bhabha National Institute, Kolkata 700064, India}

\author{Kalpataru Pradhan}
\affiliation{Theory Division, Saha Institute of Nuclear Physics,
A CI of Homi Bhabha National Institute, Kolkata 700064, India}

\date{\today}

\begin{abstract}
We investigate the impact of random bond disorder on quantum skyrmions using a spin-half quantum Heisenberg model on the square lattice with 
Dzyaloshinskii-Moriya interaction, Heisenberg anisotropy, and boundary-pinned magnetic field.
Utilizing the neural network quantum state technique, we explore the influence of disorder on spin textures, topological properties, and quantum entanglement.
We show that the disorder reduces the stability of quantum skyrmions, ultimately causing them to collapse at high disorder strengths.
In addition, our results reveal two key insights. First, the presence of disorder, rather than simply degrading skyrmion order, significantly enhances local quantum entanglement, as evidenced by
the rise in second Rényi entropy. Second, our calculations show that the topological entanglement entropy calculated using the second Rényi entropy 
remains negligible across all the 
disorder strengths. This suggests long-range entanglement is absent and the skyrmion phase is not detectable using this specific probe.
Overall, our work provides new insights into how disorder constructively influences quantum materials.
\end{abstract}

\maketitle

\section{Introduction}
Magnetic skyrmions are topologically nontrivial chiral spin textures \cite{article,Nagaosa2013} that emerge in certain classes
of magnetic materials due to the interplay between exchange interaction, anisotropy, and 
Dzyaloshinskii–Moriya interaction (DMI)~\cite{PhysRevX.9.041063, PhysRevB.103.L060404, PhysRevB.108.094410}. 
These magnetic quasiparticle \cite{Kolech23} are of great interest for potential applications in 
spintronic devices~\cite{PhysRevLett.132.136801,Sharma2024SkyrmionsAR}, neuromorphic computing \cite{Joy2023ExtremelyEM,Lone2024AnomalousHA}, and 
memory storage \cite{Moore2024MagneticDW,He2023AllElectrical9S}
due to their key properties--nanoscale size, topological protection, and the ability to be manipulated with low energy~\cite{PhysRevB.110.155152, PhysRevResearch.4.023111}. 

Classically, skyrmions are topologically protected magnetic structures. The crucial role of several parameters in shaping skyrmion properties has been a major focus of research. 
For example, their stabilization and dynamics depend on the DMI~\cite{Srivastava2018LargeVoltageTO,Chhabra2023ManipulationOH}, particularly on its symmetry and strength. Furthermore, magnetic anisotropy 
strongly affects their formation and stability~\cite{deAssis2023SkyrmionMI,Sara2024VoltagecontrolledMA,Yang2024StraindrivenSS}. Collectively, these works clearly highlight the delicate balance of competing interactions essential for sustaining these skyrmionic textures. 
Overall, classical skyrmions have been extensively studied both theoretically and experimentally~\cite{PhysRevResearch.2.033075,Raimondo2022TemperatureGradientDrivenMS,Yambe2024DynamicalGO,Tokura2020MagneticSM,Zhang2020RobustST,Pllath2017DynamicalDI,Zheng2017DirectIO,Juge2019CurrentDrivenSD}, but their robustness to different perturbations is still being thoroughly examined~\cite{Brearton2020MagneticSI,Zhang2020RobustST}.

In nanoscale skyrmions, quantum effects become a crucial factor, necessitating a quantum treatment of the constituent spins. Quantum skyrmions are now being investigated to
analyze the relationship between topological order and quantum effects~\cite{PhysRevResearch.4.023111,PhysRevB.110.104411,PhysRevX.9.041063,PhysRevB.108.094410,Sotnikov2020ProbingTT}. However, current works on these systems are largely limited to idealized, perfectly ordered materials.
In realistic materials, imperfections are unavoidable: quenched disorder can arise from lattice distortions, 
random variations in exchange couplings, or impurities~\cite{Deng2024JammingIA,Lubomirsky2023QuenchedDA}. 
Such disorder modifies the underlying interactions and thus reshapes the stability of topological states~\cite{Liu2020QuantumcriticalSP,Shtanko2018StabilityOP}. 
Traditionally, disorder has been regarded as a destructive element that breaks symmetries, localizes states, and ultimately degrades ordered phases~\cite{Dey2020RandombondDI,Yamaguchi2019VarietyOO}.
The interplay between disorder and topology is far from trivial. 
Indeed, much of the foundational theory of topological phases assumes a clean system~\cite{Vaidya2023TopologicalPO,Zhuo2023TopologicalPI}.
However, recent discoveries have begun to paint a more nuanced picture, suggesting that 
disorder can also play a constructive role, for example, by stabilizing novel phases, enhancing quantum fluctuations, 
or reorganizing quantum correlations in unexpected ways~\cite{Rangi2024OutOT,Bera2024DisorderinducedDA,Chioquetta2024StabilityOQ}. 
This paradigm shift necessitates a systematic investigation into how disorder influences complex quantum phases like quantum skyrmions,
moving beyond the simple assumption of degradation and into a detailed exploration of its effects on quantum nature.

In this work,
we present a comprehensive numerical study of quantum skyrmions under the influence of random bond disorder. 
Our model is a spin-half quantum Heisenberg Hamiltonian on a square lattice, chosen for its relevance to two-dimensional magnetic 
materials. We include DMI, which stabilizes the chiral character of skyrmions; Heisenberg anisotropy, which helps stabilize the skyrmion spin texture; and
a boundary-pinned magnetic field that mimics a ferromagnetic embedding ~\cite{PhysRevX.9.041063} and suppresses edge effects, thereby stabilizing the skyrmion spin texture in the finite lattice.~\ref{sec:Model}.
Because disordered quantum many-body systems 
are computationally challenging, conventional approaches such as exact diagonalization become intractable for large system 
sizes. To overcome this limitation, we employ the Neural Network Quantum States (NQS) framework~\cite{Science.355.602,Carleo_2017,PhysRevB.110.104411},
implemented within the NetKet framework~\cite{netket3:2022,netket2:2019}, which offers a flexible variational representation of the wave function. 
This powerful variational 
method \ref{sec:Methods} can efficiently approximate ground states of large quantum systems, enabling us to investigate the interplay of 
disorder, entanglement, and topology.

Our investigation focuses on quantifying the impact of disorder on several key properties of 
the system. First, we examine the stability~\ref{sec:stability} of the quantum skyrmion by analyzing its spin 
texture and local spin expectation values across the lattice, then we analyze its stability, quantitatively
by computing two topological indicators, \(C\) and \(Q\), 
which capture integer winding numbers and stability-dependent properties, respectively~\cite{PhysRevB.103.L060404,Berg:1981er} 
We find that increasing disorder distorts the skyrmion profile and ultimately 
causes them to collapse, in line with the intuitive expectation that randomness acts to degrade ordered phases. 

However, at the same time, our 
calculations reveal two crucial insights. First, we find that disorder substantially enhances local quantum entanglement, as measured by the second 
Rényi entropy~\cite{PhysRevLett.104.157201,Becca_Sorella_2017}, estimated from Monte Carlo samples of the NQS wave function \ref{sec:entanglement}, 
This counterintuitive result suggests that disorder can act to distribute entanglement across the system.
Second, our analysis of topological entanglement entropy (TEE)~\cite{PhysRevLett.96.110404,mauron2024predictingtopologicalentanglemententropy}, 
computed via the Kitaev-Preskill construction using Rényi-2 entropy \ref{sec:TEE}, is found 
to be negligible at all disorder strengths, suggesting the absence of long-range entanglement. 
While previous work has shown that TEE derived from the von Neumann entropy can identify skyrmion phase on a triangular lattice~\cite{vijayan2023topologicalentanglemententropyidentify}, 
we find that an analogous procedure with Rényi-2 entropy is not sensitive to probing the skyrmion phase in our system on a square lattice. 

Finally, we explore how varying microscopic parameters modify disorder effects. 
We systematically tune the DMI strength and the anisotropy parameter \ref{sec:varDA}, 
and analyze their influence on stability and entanglement growth. 
Our results provide a detailed picture of how quantum skyrmions respond to random bond disorder, 
revealing both the fragility of their topological texture and the surprising enhancement of local quantum entanglement. 
These findings offer guidance for the design of robust devices that aim to exploit quantum skyrmions as quasiparticles, 
and underscore the broader role of disorder as a parameter that can reshape the quantum landscape rather than merely degrading it. 

\section{Model Hamiltonian and Methods}\label{sec:Model}\label{sec:Methods}
We consider a spin-half quantum Heisenberg model on a two-dimensional
square lattice without any periodic boundary conditions. The system includes DMI, Heisenberg anisotropy, and
subjected to a strong external magnetic field applied only at the lattice boundary.
The Hamiltonian of the system is given by~\cite{PhysRevB.108.094410,PhysRevResearch.4.023111}:
\begin{equation}
\begin{aligned}
H ={} & -J \sum_{\langle ij \rangle} \left( \sigma^x_i \sigma^x_j + \sigma^y_i \sigma^y_j \right)
      - A \sum_{\langle ij \rangle} \sigma^z_i \sigma^z_j \\
     & - D \sum_{\langle ij \rangle} \left( \mathbf{r}_{ij} \times \hat{z} \right) \cdot \left( \boldsymbol{\sigma}_i \times \boldsymbol{\sigma}_j \right)
      + B_z \sum_{i} \sigma^z_i
\end{aligned}
\end{equation}
where, $\sigma_i^x$, $\sigma_i^y$, and $\sigma_i^z$ represent the Pauli spin matrices corresponding to
the $x$-, $y$-, and $z$-components of the spin operator $\boldsymbol{\sigma}_i$ at lattice site $i$. The parameter $J$ denotes 
the exchange interaction strength, $A$ is the anisotropy parameter, $D$ represents the 
DMI strength, and $B_z$ is the external magnetic field applied only at 
the boundary of the square lattice to fix the spins. The vector $\mathbf{r}_{ij}$ is a unit vector 
pointing from site $i$ to site $j$, and $\hat{z}$ is the unit vector in the $z$-direction. The cross 
product $\mathbf{r}_{ij} \times \hat{z}$ determines the orientation of DMI, leading to the formation of Néel-type skyrmions. In our calculations, 
we set $J = 1$ and measure all the parameters in terms of $J$. We use $A = 0.3$, $D = 0.8$ and $B_z = 10$ unless otherwise specified.

Now to introduce random bond disorder, we replace the uniform coupling constants $J$, $A$, and $D$ with
bond-dependent terms $J_{ij}$, $A_{ij}$, and $D_{ij}$, defined as~\cite{PhysRevB.110.155152}:
\begin{align}
J_{ij} &= J \left( 1 + \rho_{ij} \, \delta \right), \\
A_{ij} &= A \left( 1 + \rho_{ij} \, \delta \right), \\
D_{ij} &= D \left( 1 + \rho_{ij} \, \delta \right),
\end{align}
where $\rho_{ij}$ is a dimensionless random variable sampled from a Gaussian distribution with zero mean and unit variance. The parameter $\delta$ controls
the strength of the disorder. We use independent Gaussian distributions for each type of coupling, and once 
the values of $\rho_{ij}$ are assigned, they remain fixed throughout the simulation for a given disorder strength. 
This approach reflects a more realistic description of physical disorder compared to sampling $\rho_{ij}$ from uniform distributions, where $\rho_{ij}$ takes values from 
$[-\delta, \delta]$, or from discrete distributions, where $\rho_{ij}$ is drawn from $\{-\delta, \delta\}$. These alternative distributions are briefly considered in Sec.~\ref{sec:entanglement}.

For all the measured quantities, the results were averaged over 50 independent disorder realizations unless otherwise specified. 
Throughout this study, a single disorder strength $\delta$ was applied uniformly to all coupling
terms. The case where each interaction is assigned an independent disorder strength is
briefly considered in Sec.~\ref{sec:entanglement}.

\begin{figure}[htb]
    \centering
    \includegraphics[width=0.72\linewidth]{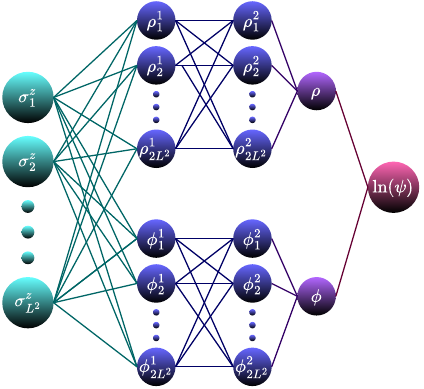}
    \caption{Architecture of the NQS. The input is a lattice spin configuration in the $\sigma^z$ basis. Two separate neural networks are used to model the modulus $\rho$ and 
    phase $\phi$ with each layer consists of \(2L^2\) neurons. The output is the logarithm of the wave function.}
    \label{fig:nqs}
\end{figure}
Now, to find the ground state wave function
of quantum skyrmions we used the NQS approach, similar to that described in \cite{PhysRevB.108.094410}. 
The variational wave function is given by:
\begin{equation}
|\psi_\theta\rangle = \sum_{\boldsymbol{\sigma}} \psi_\theta(\boldsymbol{\sigma}) |\boldsymbol{\sigma}\rangle,
\end{equation}
where \( |\boldsymbol{\sigma}\rangle \) denotes the computational basis, which in our case is chosen as eigenstates of the \( \sigma^z \) operator, 
and \( \theta \) represent the set of trainable parameters of the network.
\begin{figure*}[htb]
  \centering
  \includegraphics[width=0.68\columnwidth]{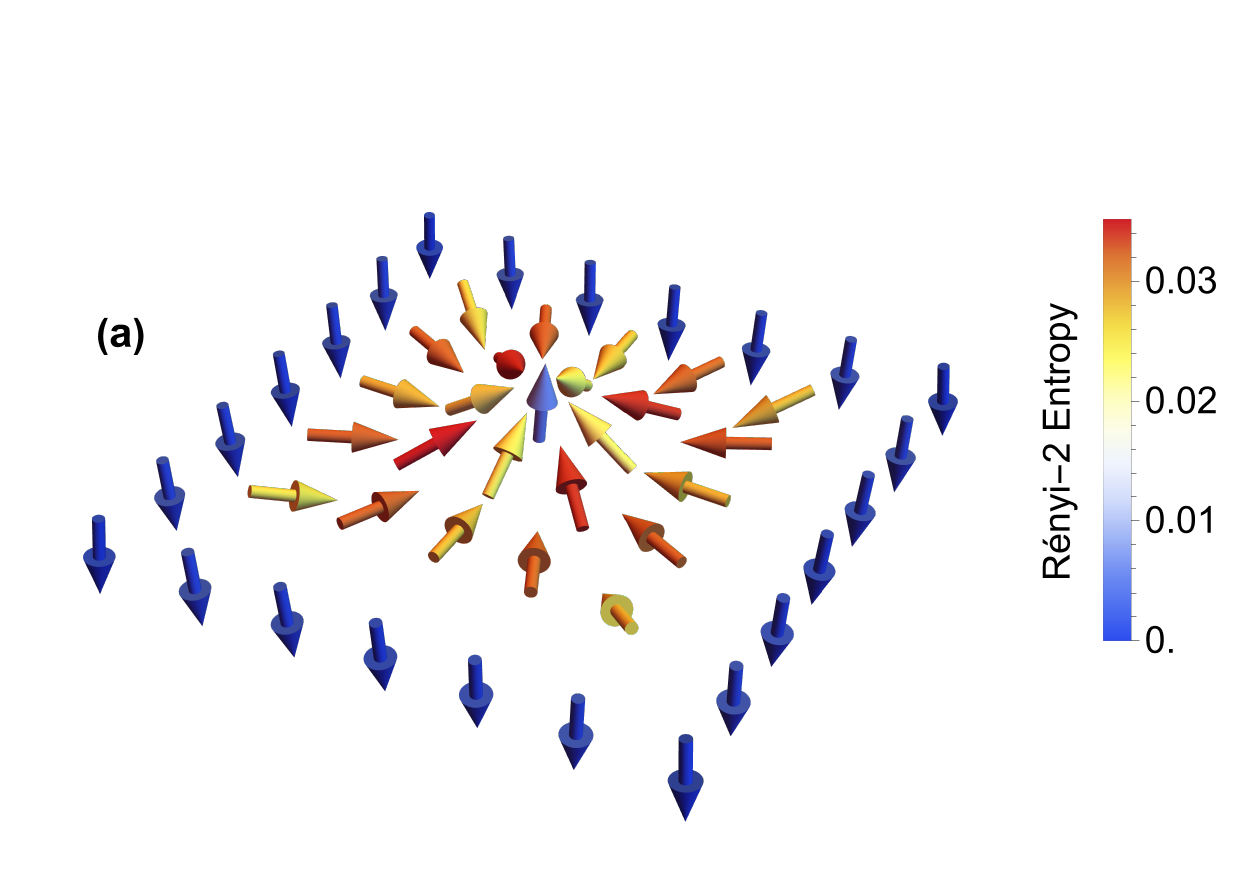}
  \includegraphics[width=0.68\columnwidth]{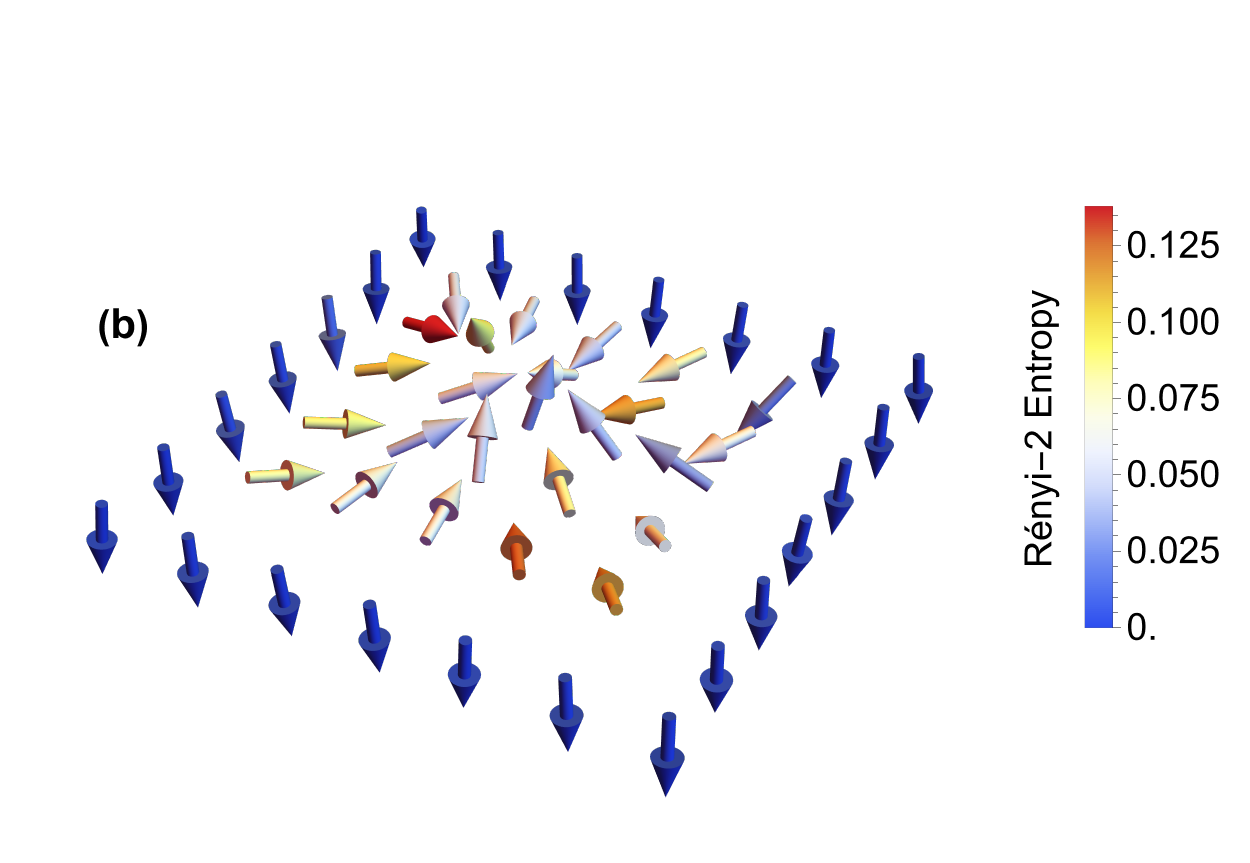}
  \includegraphics[width=0.68\columnwidth]{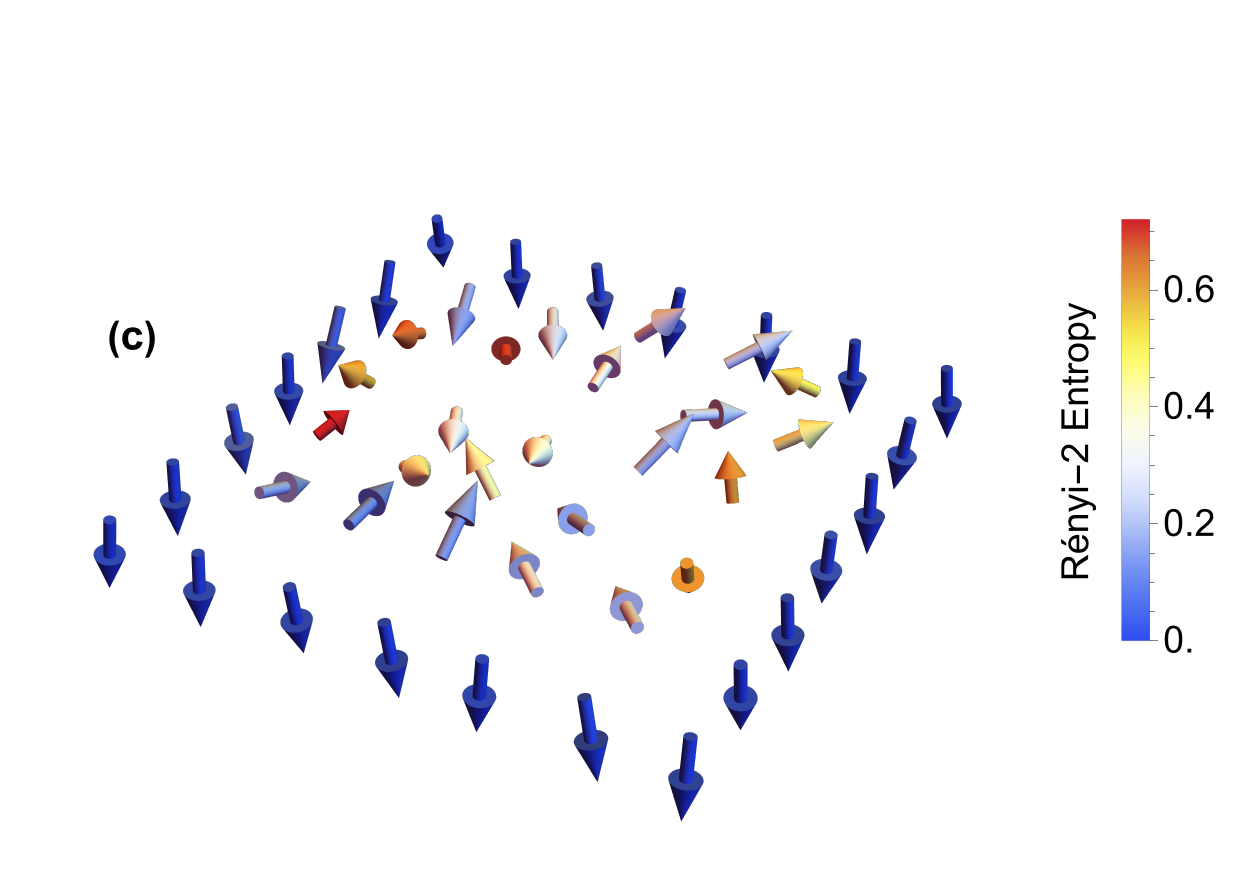}\\
  \includegraphics[width=0.68\columnwidth]{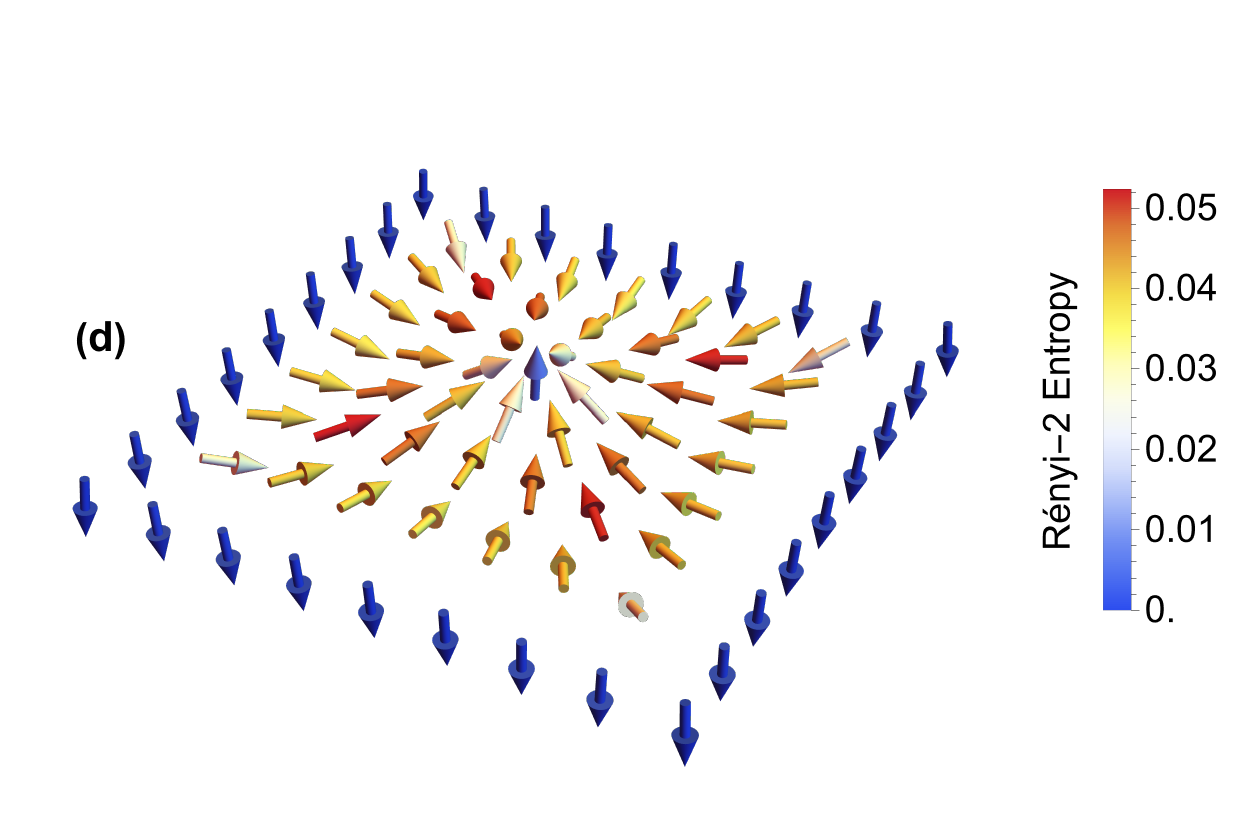}
  \includegraphics[width=0.68\columnwidth]{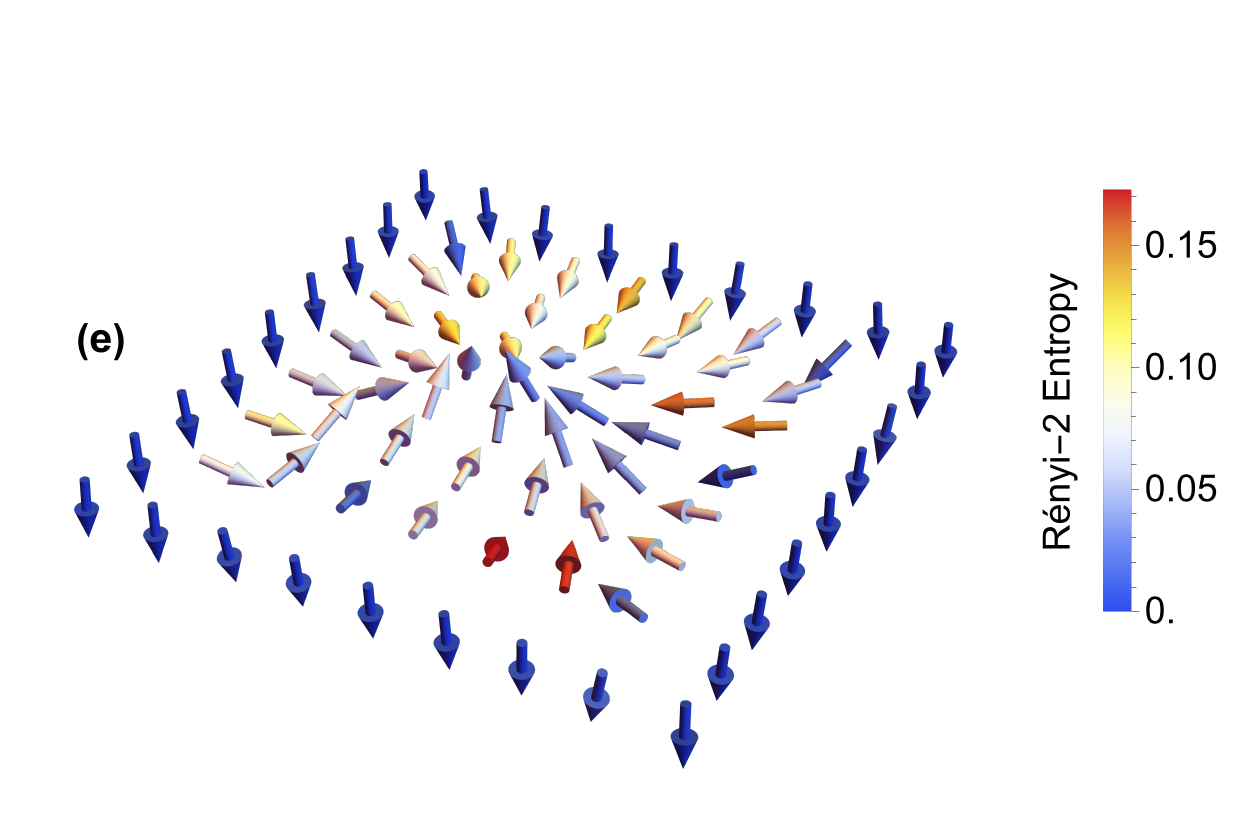}
  \includegraphics[width=0.68\columnwidth]{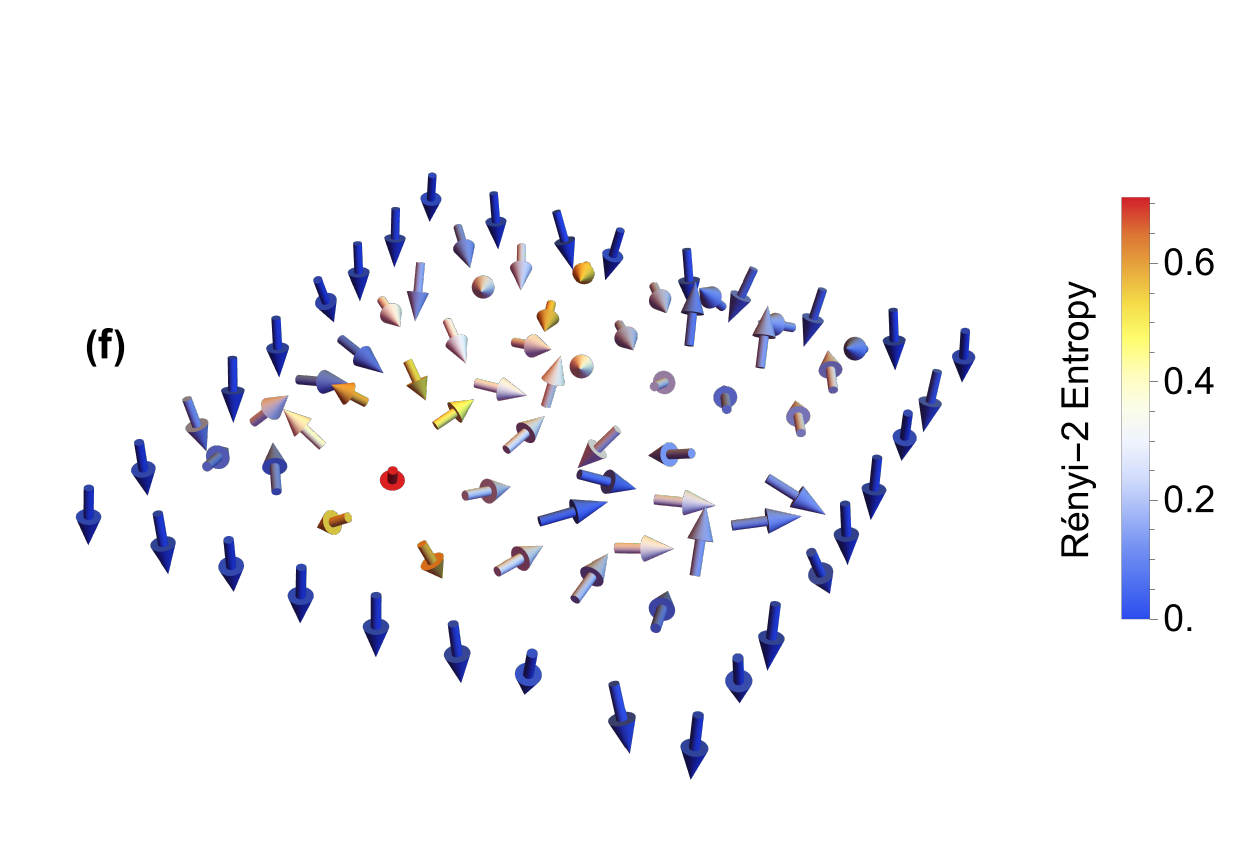}
  \caption{Spin configurations for \( L = 7 \) (a--c) and \( L = 9 \) (d--f) at different disorder strengths \( \delta \) for \(D = 0.8\) and \(A = 0.3\), with color 
  indicating local Rényi-2 entropy, and arrow length indicates the magnitude of the spin expectation value. Panels show (a,d) \( \delta = 0.0 \); (b,e) \( \delta = 0.5 \); and 
  (c,f) \( \delta = 1.5 \).}
  \label{fig:skyrmions}
\end{figure*}
The real and imaginary parts of the logarithm of the wave function are represented by two separate fully connected feedforward neural networks. The logarithm of the wave function is given by:
\begin{equation}
\ln \psi_\theta(\boldsymbol{\sigma}) = \rho(\boldsymbol{\sigma}) + i \phi(\boldsymbol{\sigma}),
\end{equation}
where \( \rho \) and \( \phi \) are the outputs of the respective networks modeling the modulus and the phase of the wave function.
The spin configuration of the lattice is used as an input. Each neural network (for \( \rho \) and \( \phi \)) consists of 
two fully connected hidden layers, each having \( 2 L^2 \) neurons (see Fig.~\ref{fig:nqs}), where \( L \) is the size of our square lattice. 
We used ReLU (Rectified Linear Unit) as the activation function throughout the hidden layers.

Training of the variational parameters \( \theta \) is performed by minimizing the expectation value of the Hamiltonian:
\begin{equation}
\mathcal{L}_\theta = \langle \psi_\theta | \hat{H} | \psi_\theta \rangle,
\end{equation}
which acted as the loss function. We initialize training by optimizing the phase part of the network 
first, keeping the modulus part fixed as it helps the network to learn the sign structure more effectively \cite{PhysRevResearch.2.033075}. 
To optimize the parameters, we employ the Adam optimizer. Sampling of spin configurations is 
carried out using the Markov Chain Monte Carlo technique. The overall implementation, including both the 
neural network architecture and the sampling procedure, is done using the open-source NetKet library
\cite{netket3:2022,netket2:2019,mpi4jax:2021}.
Further implementation and training details are provided in Appendix~\ref{app:NQS}.

\section{Spin Texture and Stability in Presence of Disorder}\label{sec:stability}

To study how disorder affects the spin configurations of quantum skyrmions, we choose \( D = 0.8 \) and \( A = 0.3 \)
for which the ground state corresponds to a stable skyrmion at $\delta = 0.0$, which is further quantified, later in this section. Fig.~\ref{fig:skyrmions}(a--c) shows the spin textures for different disorder strengths 
for a lattice of size \( L = 7 \). The arrow length represents the spin magnitude, while the color denotes the Rényi-2 entropy (discussed later in Sec.~\ref{sec:entanglement}).  
At \(\delta = 0.0\), we obtain a well-defined single skyrmion spanning the entire lattice.
Moderate disorder strength (\(\delta = 0.5\)) slightly perturbs the central spin, but the skyrmion texture persists.
However, at high disorder strength (\(\delta = 1.5\)), the skyrmion is completely destabilized, and the spin magnitude indicated by arrow length decreases at several lattice sites.  

Now to compare how the lattice size influences the spin configurations, we also consider the case of \( L = 9 \), as shown in Fig.~\ref{fig:skyrmions}(d--f).
At \(\delta = 0.0\), a single skyrmion is again observed.  
For \(\delta = 0.5\), again the central spin is slightly tilted, but the texture retains the skyrmion character.
Finally, at \(\delta = 1.5\), the skyrmion is fully destabilized and the overall spin magnitude is reduced at many lattice sites.
In both cases, the qualitative nature of the effect of disorder remains the same for $L = 7$ and $L = 9$.

\begin{figure}[htb]
    \centering
    \includegraphics[width=0.49\linewidth]{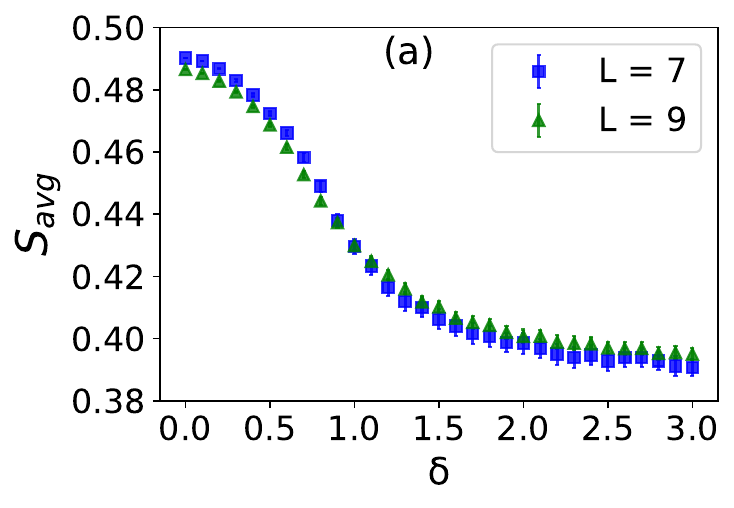}
    \includegraphics[width=0.49\linewidth]{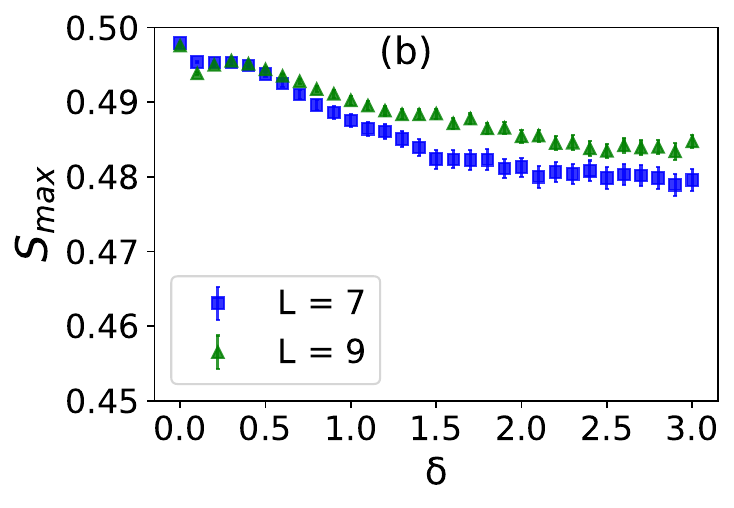}\\[1ex]
    \includegraphics[width=0.49\linewidth]{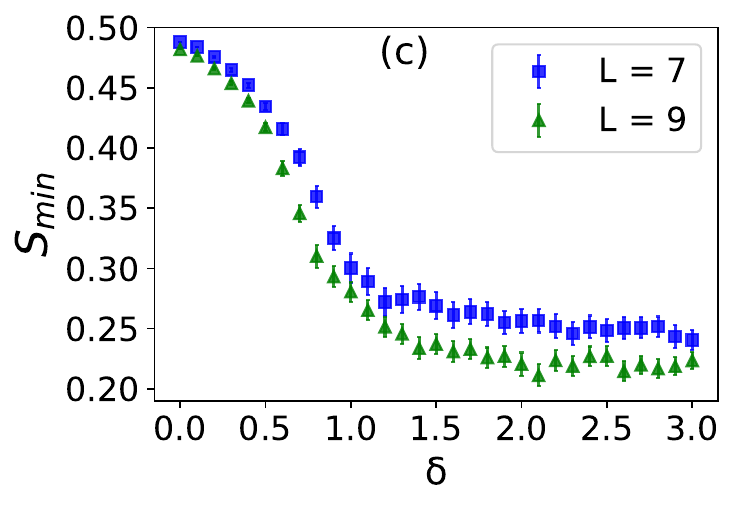}
    \includegraphics[width=0.49\linewidth]{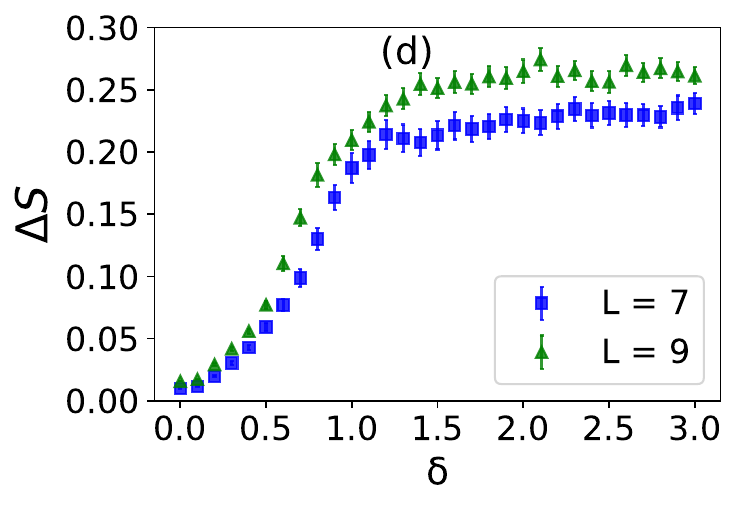}
    \caption{(a) Average spin magnitude, (b) Maximum spin magnitude, (c) Minimum spin magnitude, and (d) Spin 
    magnitude range as a function of disorder strength for \( L = 7 \) and \( L = 9 \), with standard error bars over 50 disorder realizations.}
    \label{fig:spin_expectation}
\end{figure}
To make the analysis of the spin structure in the presence of disorder clearer and more quantitative,
we evaluate several quantities from the local 
spin expectation values at each lattice site \(i\), excluding sites at the boundary of the lattice. The magnitude of the spin 
expectation value at site \(i\) is defined as:
\begin{equation}
|\langle \vec{S}_i \rangle| = \sqrt{ \langle S_i^x \rangle^2 + \langle S_i^y \rangle^2 + \langle S_i^z \rangle^2 }.
\label{eq:Spin_magnitude}
\end{equation}
Using $|\langle \vec{S}_i \rangle|$, we compute the average spin magnitude ($S_{\text{avg}} = L^{-2} \sum_i |\langle \vec{S}_i \rangle|$).
We extract the maximum spin magnitude ($S_{\text{max}} = \max_i |\langle \vec{S}_i \rangle|$), the minimum spin magnitude ($S_{\text{min}} = \min_i |\langle \vec{S}_i \rangle|$), 
and calculate the spin magnitude range ($\Delta S = S_{\text{max}} - S_{\text{min}}$).
These quantities help to identify variations of the spin expectation across the system under disorder.

Now, to study how these quantities vary with disorder strength, we have taken the average of each quantity
over 50 disorder realizations. The corresponding errors are calculated as the standard error of the mean,
and the results are plotted in Fig.~\ref{fig:spin_expectation}. For \( L = 7 \),
\( S_{\text{avg}} \) [Fig.~\ref{fig:spin_expectation}(a)] decreases with increasing disorder strength.
A sharp reduction begins around \( \delta = 0.7 \), which corresponds to the regime where the quantum skyrmion
starts to degrade, beyond which \( S_{\text{avg}} \) tends to saturate,
in the regime where the skyrmion is completely degraded, which is discussed in detail later in this section.
\( S_{\text{max}} \), shown in Fig.~\ref{fig:spin_expectation}(b), exhibits only a very slight decrease with disorder
and remains largely unaffected throughout the range of \( \delta \). In contrast,
\( S_{\text{min}} \), shown in Fig.~\ref{fig:spin_expectation}(c), decreases more strongly with disorder than both
\( S_{\text{avg}} \) and \( S_{\text{max}} \), and follows a similar trend, with a sharp drop around \( \delta = 0.7 \) that continues until
\( \delta = 1.0 \), after which it saturates. The difference \( \Delta S = S_{\text{max}} - S_{\text{min}} \),
shown in Fig.~\ref{fig:spin_expectation}(d), increases with disorder strength, showing a pronounced rise near
\( \delta = 0.7 \) and saturate beyond \( \delta = 1.0 \). This behavior reflects the fact
that the variation in \( \Delta S \) is mainly governed by the strong suppression of \( S_{\text{min}} \),
as \( S_{\text{max}} \) remains nearly constant with increasing disorder.
To compare our results for a larger system size,
we have also plotted the corresponding data for \( L = 9 \) in each panel of Fig.~\ref{fig:spin_expectation} together with
the \( L = 7 \) data, and we find that the trends for \( L = 9 \) are qualitatively similar to those observed for \( L = 7 \).

Using the local spin expectation values $\langle \vec{S}_i \rangle$, we computed two related quantities, skyrmion number \( C \) and skyrmion stability \( Q \) as described in \cite{PhysRevResearch.4.023111,PhysRevB.103.L060404}.
Both \( C \) and \( Q \) are defined using the same geometric formula based on the scalar 
triple product of spins on triangular plaquettes:
\begin{equation}
C, Q = \frac{1}{2\pi} \sum_{\Delta} \tan^{-1} \left( 
\frac{ \vec{n}_i \cdot (\vec{n}_j \times \vec{n}_k) }
{ 1 + \vec{n}_i \cdot \vec{n}_j + \vec{n}_j \cdot \vec{n}_k + \vec{n}_k \cdot \vec{n}_i } 
\right),
\label{eq:CQ}
\end{equation}
where the sum runs over all elementary triangles \( \Delta \) on a triangulated square lattice, 
and the indices \( i, j, k \) refer to the lattice sites corresponding to three vertices of each triangle.
The difference between \( C \) and \( Q \) lies in the definition of the local 
spin vector \( \vec{n}_i \), which is constructed from the expectation value of the spin operator 
at each site
(\(\langle \vec{S}_i \rangle = (\langle S_i^x \rangle, \langle S_i^y \rangle, \langle S_i^z \rangle)^T\)).
The definition of $\vec{n}_i$ used for computing $C$ is $\langle \vec{S}_i \rangle / |\langle \vec{S}_i \rangle|$, while for $Q$ it is $2 \langle \vec{S}_i \rangle$.
Here, $|\langle \vec{S}_i \rangle|$ is computed using Eq.~\ref{eq:Spin_magnitude}.
The quantity \( C \) is computed using normalized $\vec{n}_i$, which counts only the angular winding of the spin configuration irrespective of the spin magnitude, and therefore takes only integer values.
In contrast, \( Q \) is computed using unnormalized $\vec{n}_i$, which
includes the effect of spin magnitude and hence can take non-integer values.

Now, in disordered systems, for a single realization of disorder, the value of any observable will fluctuate as the disorder strength increases. 
The same behavior is observed for both \( C \) and \( Q \); however, since \( C \) is defined to take only integer values, it can fluctuate only between integer values. 
As mentioned earlier, for disordered systems, the physically meaningful quantities are obtained by averaging over many disorder realizations. 
Therefore, we average both \( C \) and \( Q \) over 50 disorder realizations.
As a result, \( C \) can take non-integer values in the averaged data.
Now, to determine whether the quantum skyrmion can exist in the ground state, 
we take into account the $C$.
In our averaged data, we obtain skyrmions only for \( C = 1 \); any deviation from unity or a non-integer value of averaged \( C \) does not correspond to a skyrmion spanning the entire lattice.
To further analyze the stability of these skyrmions when \( C = 1 \), we consider the quantity \( Q \). 
For \( Q = 1 \), we obtain the most stable skyrmion state. 
As the value of \( Q \) decreases, the stability of the skyrmion against local perturbations also decreases. 
This reduction occurs because the value of \( Q \) decreases as spin expectation value \(\langle \vec{S}_i \rangle\) becomes smaller, 
due to which, small changes in \(\langle \vec{S}_i \rangle\) can lead to spin flips and alter the corresponding topological index~\cite{PhysRevResearch.4.023111,PhysRevB.103.L060404}.

\begin{figure}[htb]
    \centering
    \includegraphics[width=0.49\linewidth]{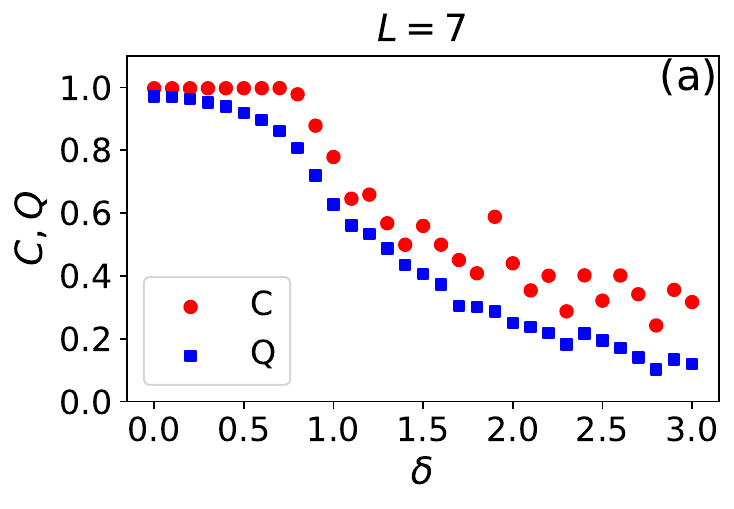}
    \includegraphics[width=0.49\linewidth]{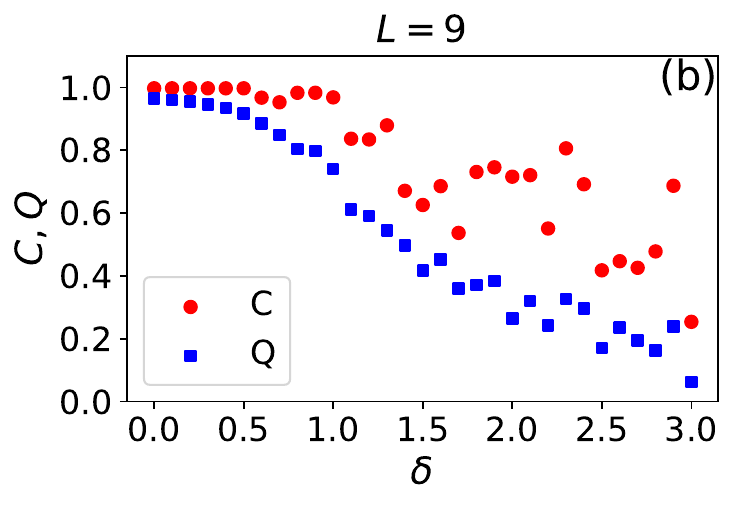}
    \caption{Disorder-averaged skyrmion number \( C \) and skyrmion stability \( Q \) as a function of disorder strength for
    (a) \( L = 7\) and (b) \( L = 9\), at fixed \( D = 0.8 \) and \( A = 0.3 \).}
    \label{fig:c_and_ql79}
\end{figure}

Now to quantitatively examine how disorder impacts quantum skyrmions, we compute the disorder-averaged values of \( C \) and \( Q \) as functions of disorder strength \( \delta \) 
for lattice size \( L = 7 \) at \( D = 0.8 \) and \( A = 0.3 \) as shown in Fig.~\ref{fig:c_and_ql79}(a).  
At zero disorder (\( \delta = 0 \)), the ground state corresponds to a single stable skyrmion with \( C = 1 \) and \( Q = 1 \).  
As the disorder increases, \( Q \) decreases, while the value of \( C \) remains unity.  
After \(\delta \approx 0.7\), \( C \) starts to deviate from unity toward zero, indicating the absence of a skyrmion spanning the entire lattice and with further increase in disorder strength, we 
obtain more instances of \( C = 0 \) in the averaged data, resulting in non-integer values of \( C \) that decrease toward zero. In this regime, \( Q \) is also very small and decreases 
toward zero, indicating a high degree of instability.
To analyze whether \( C = 1 \) for $\delta < 0.7$ corresponds to a stable skyrmion in the ground state, we consider the skyrmion stability.  
The value of \( Q \) decreases continuously with increasing disorder and drops sharply at \(\delta \approx 0.7\) to below 0.9.  
This indicates that beyond this point, skyrmions become highly unstable and disordered due to the reduced spin magnitude.

To study the effect of lattice size, we repeated the calculations for \( L = 9 \) and at same parameters, \( D = 0.8 \) and \( A = 0.3 \), which also lead to single stable skyrmion in ground state at $\delta = 0$, shown in Fig.~\ref{fig:c_and_ql79}(b).  
For \( L = 9 \), the deviation of \( C \) from unity occurs at (\(\delta \approx 0.7\)), similar to that in \( L = 7 \).  
The skyrmion stability \( Q \), again, decreases immediately after disorder is introduced, signaling reduced stability of the skyrmion phase at higher disorder strengths.
Overall, we observe that the behavior of both \( C \) and \( Q \) is similar for both lattice sizes. 
In summary, we find that disorder reduces the stability of quantum skyrmions, eventually causing them to collapse at high disorder strengths.

\section{Entanglement and Its Response to Disorder}\label{sec:entanglement}
Quantum entanglement provides a measure of correlations in many-body systems that goes beyond conventional observables such as magnetization. 
In disordered quantum systems, entanglement serves as a tool to characterize the phases, transitions, and localization properties. 
To quantitatively examine entanglement, we employ the second Rényi entropy, 
which offers both conceptual clarity and computational efficiency. The second Rényi entropy \( S_2 \) can be directly evaluated from the expectation value of a swap operator in numerical simulations, 
and it captures the scaling of entanglement with subsystem size and disorder strength. 
To calculate \( S_2 \) per lattice site, we partition the lattice into two subsystems: 
region \( A \), which consists of a single spin, and region \( B \), which includes the rest of the lattice.
The second Rényi entropy is defined as \cite{PhysRevLett.104.157201}:
\begin{equation}
S_2(\rho_A) = -\ln \left( \text{Tr}(\rho_A^2) \right) = -\ln \left( \langle \text{Swap}_A \rangle \right),
\end{equation}
where \( \langle \text{Swap}_A \rangle \) is the expectation value of the swap operator acting on region \( A \).
In practice, we estimate \( \langle \text{Swap}_A \rangle \) using Monte Carlo sampling over \( N_s \) pairs 
of spin configurations \cite{PhysRevResearch.2.023358} which is given by:
\begin{equation}
\langle \text{Swap}_A \rangle 
\approx \frac{1}{N_s} \sum_{i=1}^{N_s} 
\frac{
\psi_\theta(\tilde{\boldsymbol{\sigma}}^{(i)}_A, \boldsymbol{\sigma}^{(i)}_B) \,
\psi_\theta(\boldsymbol{\sigma}^{(i)}_A, \tilde{\boldsymbol{\sigma}}^{(i)}_B)
}{
\psi_\theta(\boldsymbol{\sigma}^{(i)}_A, \boldsymbol{\sigma}^{(i)}_B) \,
\psi_\theta(\tilde{\boldsymbol{\sigma}}^{(i)}_A, \tilde{\boldsymbol{\sigma}}^{(i)}_B)
}.
\end{equation}
where \( \boldsymbol{\sigma}^{(i)} \) and \( \tilde{\boldsymbol{\sigma}}^{(i)} \) represent 
spin configurations sampled independently from two copies of the variational wave function.

\begin{figure}[htb]
    \centering
    \includegraphics[width=0.94\linewidth]{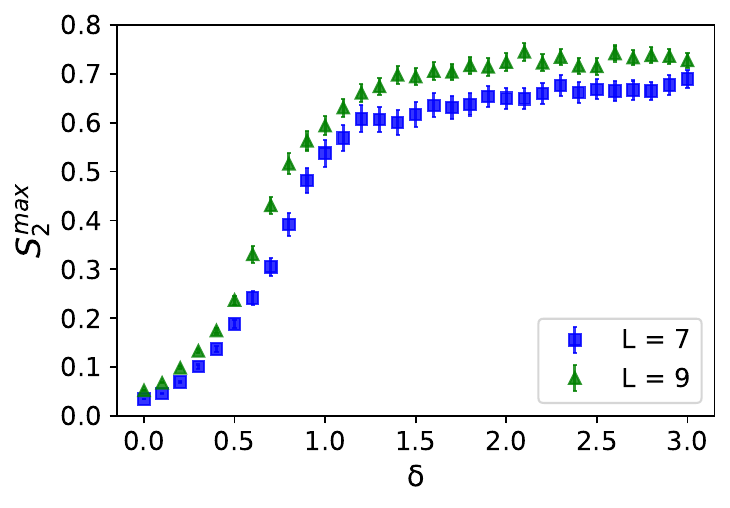}
    \caption{Maximum Rényi-2 entropy as a 
    function of disorder strength for \( L = 7 \) and \( L = 9 \) with standard error over 50 disorder realizations represented as bars.}
    \label{fig:maxRenyi}
\end{figure}

Now, to examine the effect of disorder on quantum entanglement, 
we extract the maximum value of Rényi-2 entropy for each value of \( \delta \) and average it over 50 disorder samples. 
as shown in Fig.~\ref{fig:maxRenyi}. 
For $L = 7$, at zero 
disorder, the maximum Rényi-2 entropy is close to zero. As the disorder strength increases, the maximum Rényi-2 entropy 
starts to increase and the rise is sharp in the range from \( \delta \approx 0.7 \) to \( \delta \approx 1.0 \), the regime where skyrmion degrades and ultimately collapses. After that, it begins to saturate. 
We also plot the maximum Rényi-2 entropy for \( L = 9 \) in the same figure Fig.~\ref{fig:maxRenyi}. 
Here too, the effects are more or less the same as those of $L = 7$, maximum Rényi-2 entropy first increases with the disorder strength and then saturates.
In addition it is important to note that maximum Rényi-2 entropy follows the trends similar to that of \( \Delta S = S_{\text{max}} - S_{\text{min}} \) [Fig.~\ref{fig:spin_expectation}(d)] 
This behavior is related to the fact that
decrease in \( S_{\text{max}} \) [Fig.~\ref{fig:spin_expectation}(b)] is much weaker than in \( S_{\text{min}} \) [Fig.~\ref{fig:spin_expectation}(c)],
and the smaller spin magnitudes correspond to larger Rényi-2 entropy [Fig.~\ref{fig:skyrmions}(c)] (further discussed at the end of the next paragraph) and hence $\Delta S$ follows the same trend.

To further illustrate the effect of disorder, we show representative spin configurations for 
different values of disorder strength in Fig.~\ref{fig:skyrmions}. The color of each spin denotes 
the corresponding Rényi-2 entropy, while the vector length represents the spin magnitude as defined 
in Eq.~\ref{eq:Spin_magnitude}.
For \( L = 7 \) at \( \delta = 0 \), the system exhibits weak entanglement, as indicated by the low values of the Rényi-2 entropy.
Maximum Rényi-2 entropy is observed between the central spin and the boundary spins. Upon increasing disorder strength to \( \delta = 0.5 \), the maximum value increases,
and the entanglement pattern modifies from its $\delta = 0$ structure to a more random structure. 
Upon further 
increase in disorder strength to \( \delta = 1.5 \), the maximum Rényi-2 entropy rises even more, and the spin of lowest magnitude corresponds to the maximum Rényi-2 entropy (shown in red) becomes more pronounced.
For \( L = 9 \), as shown in Fig.~\ref{fig:skyrmions}(d--f),
the systematic behavior of Rényi-2 entropy remains more or less the same, the entanglement pattern is modified and becomes more random when disorder is increased. 
Moreover, the trend that smaller spin magnitudes correspond to larger Rényi-2 entropy remains true. It is consistent with Ref.~\cite{qxtm-gysy}, which reports that quantum entanglement degrades as the spin magnitude 
approaches the classical limit. In our case, this translates to a decrease in spin magnitude being associated with an increase in entanglement, 
because the classical limit for our system is given by \( |\langle \vec{S}_i \rangle| = 0.5 \), whereas in the quantum regime \( |\langle \vec{S}_i \rangle| \)
decreases from 0.5 as entanglement grows.

\begin{figure}[htb]
    \centering
    \includegraphics[width=0.7\linewidth]{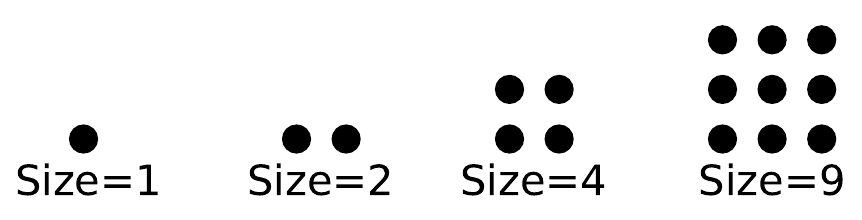}
    \includegraphics[width=0.94\linewidth]{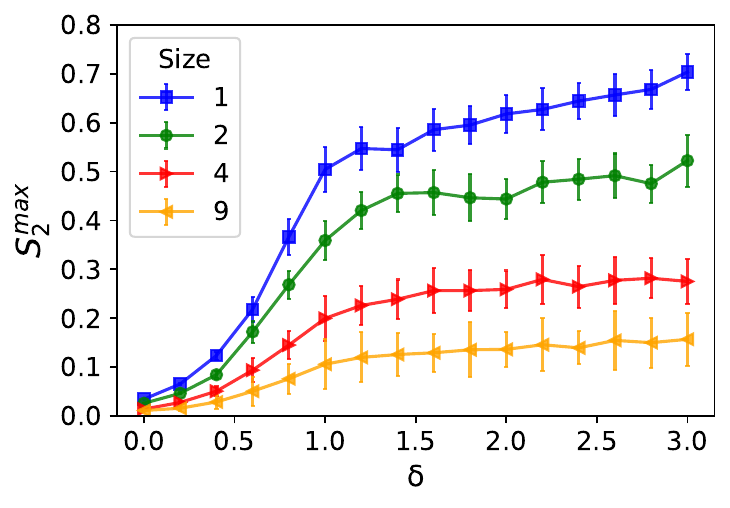}
    \caption{Maximum Rényi-2 entropy for \( L = 7 \) as a 
    function of disorder strength for different subsystem (region $A$) sizes. Error bars denote the standard error over 15 disorder realizations.}
    \label{fig:maxRenyibatch}
\end{figure}

Now to examine the nature of entanglement and its spatial extent, we study how it varies with the size of the subsystem (region $A$). 
The size of region $A$ is increased by including a larger number of neighboring spins (shown on the top of Fig.~\ref{fig:maxRenyibatch}). 
We consider all the combinations of a given subsystem spanning the entire lattice, excluding the spins at the lattice boundaries, for \( L = 7 \). We compute the maximum Rényi-2 entropy normalized according to the size of the subsystem and average the results over 15 disorder realizations. 
The results are shown in Fig.~\ref{fig:maxRenyibatch}. 
As the size of the region $A$ increases, the maximum Rényi-2 entropy decreases and saturates at smaller values. 
This behavior indicates that spins are more strongly entangled with their nearest neighbors than with distant ones, and suggests that long-range entanglement is absent in our system. 
A systematic analysis of long-range entanglement is presented in Sec.~\ref{sec:TEE}.

\begin{figure}[htb]
    \centering
    \includegraphics[width=0.49\linewidth]{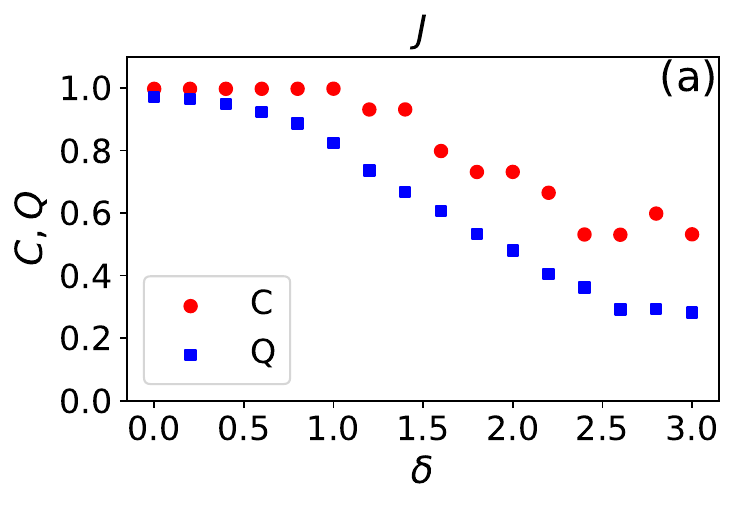}
    \includegraphics[width=0.49\linewidth]{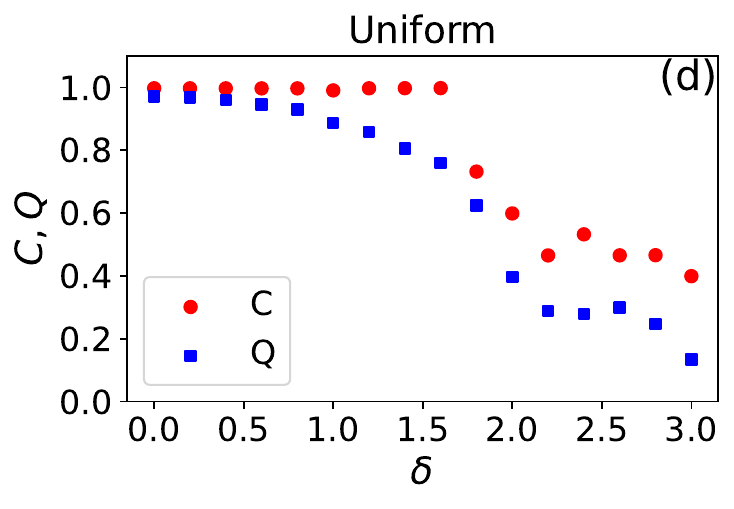}
    \includegraphics[width=0.49\linewidth]{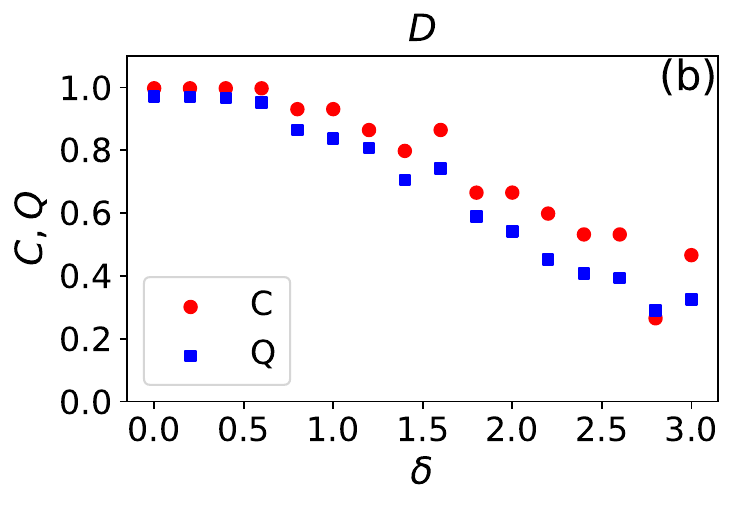}
    \includegraphics[width=0.49\linewidth]{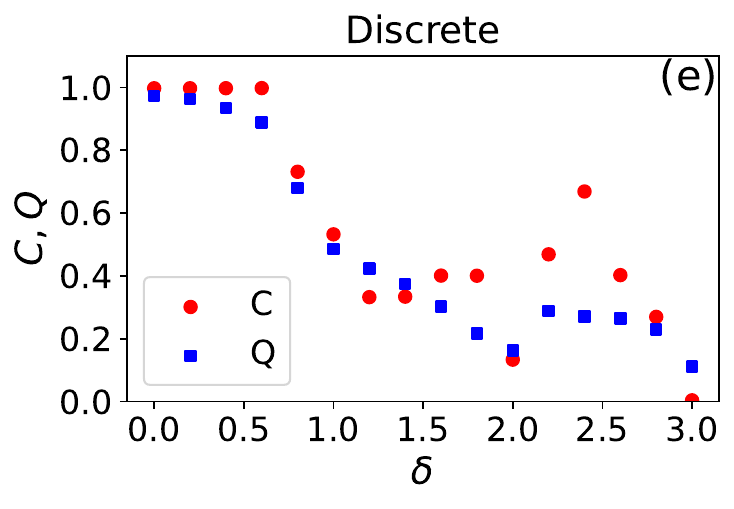}
    \includegraphics[width=0.49\linewidth]{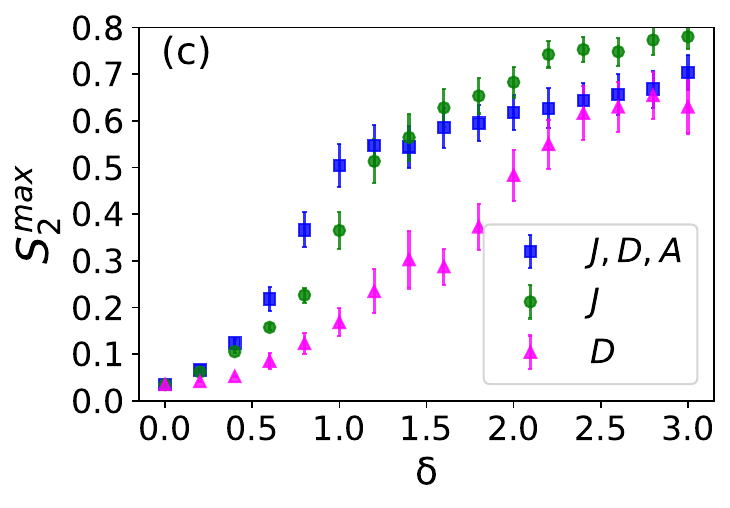}
    \includegraphics[width=0.49\linewidth]{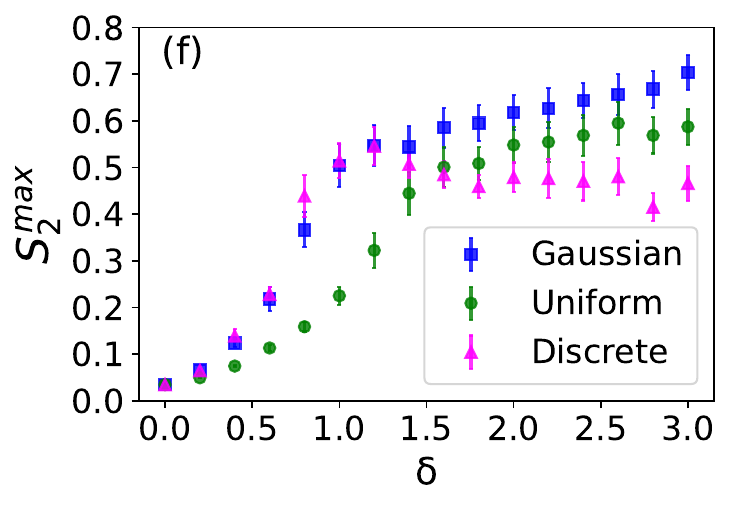}
    \caption{
    Skyrmion number $C$ and skyrmion stability $Q$ as functions of disorder strength for Gaussian disorder applied (a) only to $J$, (b) only to $D$; (c) uniform disorder, and (d) discrete disorder in all couplings. 
    Maximum Rényi-2 entropy as a function of disorder strength for (e) Gaussian disorder applied to all couplings, and separately for disorder only in $J$ and only in $D$; 
    and (f) for different types of disorder distributions: Gaussian, uniform, and discrete. 
    The error bars denote the standard error over 15 disorder realizations. The lattice size is $L = 7$.
    }
    \label{fig:maxRenyidistribution}
\end{figure}

Now, to identify whether the degradation of the skyrmion and the increase in entanglement arise solely from disorder in specific couplings or from the disorder in all couplings, 
we compute the skyrmion number and skyrmion stability for Gaussian disorder assigned only to $J$ [Fig.~\ref{fig:maxRenyidistribution}(a)] and only to $D$ [Fig.~\ref{fig:maxRenyidistribution}(b)]. 
We also compute $S_{2}^{max}$ for both disorder scenarios and compare it with the case of Gaussian disorder in all the couplings [Fig.~\ref{fig:maxRenyidistribution}(c)] on a lattice of size \( L = 7 \). 
All results are averaged over 15 disorder realizations. 
We find that skyrmions degrade in all cases; however, for disorder only in $J$, $C$ starts to deviate from unity toward zero at higher disorder strength
($\delta \approx 1.4$), in contrast to the case of disorder only in $D$, where it occurs at lower $\delta$. 
This is intuitive because skyrmions mainly arise from the interplay of DMI and Heisenberg anisotropy, so disorder in the DMI degrades them at lower disorder strength.
But for both disorder only in $J$ and only in $D$, $Q$ decreases as the disorder strength increases, suggesting instability of skyrmions at higher disorder strength.
The value of $S^{\mathrm{max}}_2$ increases in all cases, though the growth rate depends on which couplings are affected by disorder. It exhibits a sharp increase where skyrmions degrade,
which occurs earlier for disorder only in $J$ than for disorder only in $D$ following trends similar to $C$.

Now, to test the effect of different random distributions on the stability and entanglement properties of skyrmions, we also compute $C$ and $Q$ for uniform [Fig.~\ref{fig:maxRenyidistribution}(d)] and discrete disorder [Fig.~\ref{fig:maxRenyidistribution}(e)], as well as corresponding $S^{\mathrm{max}}_2$ 
and compared it with the case of Gaussian disorder [Fig.~\ref{fig:maxRenyidistribution}(f)] on a lattice of size \( L = 7 \). 
All results in Fig.~\ref{fig:maxRenyidistribution} are averaged over 15 disorder realizations. 
Here, $C$ deviates from unity toward zero at higher disorder strength, around $\delta \approx 1.6$ for the case of uniform disorder, in contrast to $\delta \approx 0.7$ for the discrete case. 
This behavior arises because the discrete distribution introduces stronger disorder, as it takes only discrete values in contrast to the continuous values of a uniform distribution. 
The quantity $Q$ decreases as the disorder strength increases, suggesting increasing instability of skyrmions at higher disorder strength. 
Similarly, $S^{\mathrm{max}}_2$ increases in all cases, though the growth rate depends on the type of distribution, and it exhibits a sharp increase where skyrmions degrade, which occurs earlier for 
discrete disorder than for the uniform one, following the trends similar to $C$.

\section{Disorder Dependence of Topological Entanglement Entropy}\label{sec:TEE}

\begin{figure}[htb]
    \centering
    \includegraphics[width=0.8\linewidth]{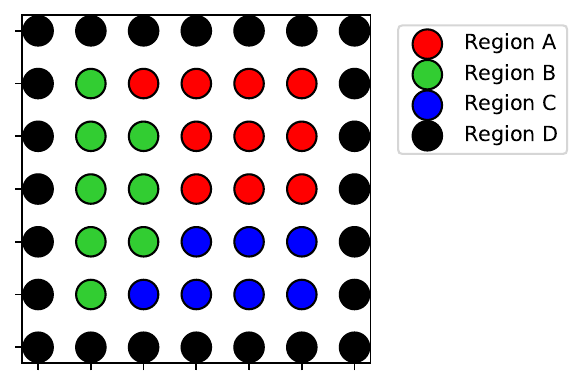}
    \caption{Schematic of the partition used for calculating the TEE for \( L=7 \). Regions $A$, $B$, 
    and $C$ are shown in red, green, and blue, respectively, while the boundary spins are
    represented by region $D$.}
    \label{fig:tee_region}
\end{figure}

\begin{figure*}[htb]  
  \centering
  \includegraphics[width=0.68\columnwidth]{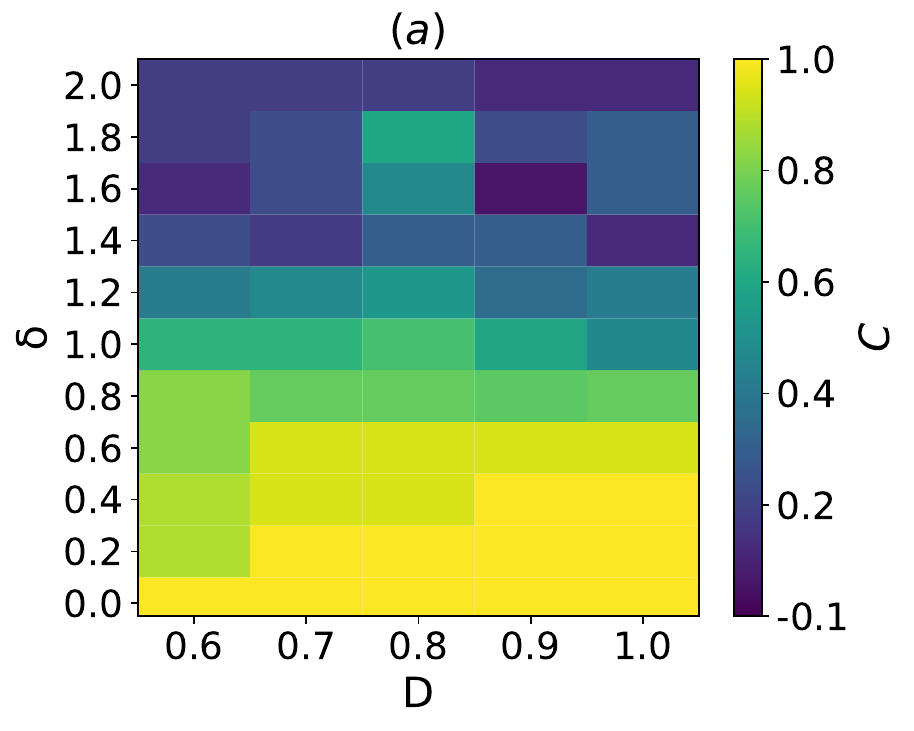}
  \includegraphics[width=0.68\columnwidth]{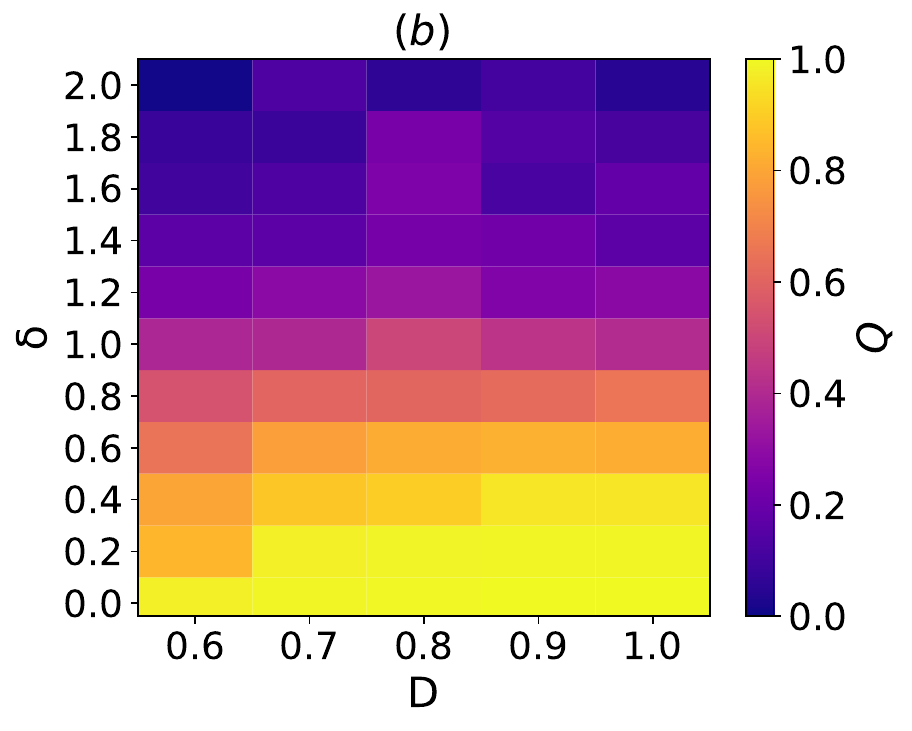}
  \includegraphics[width=0.68\columnwidth]{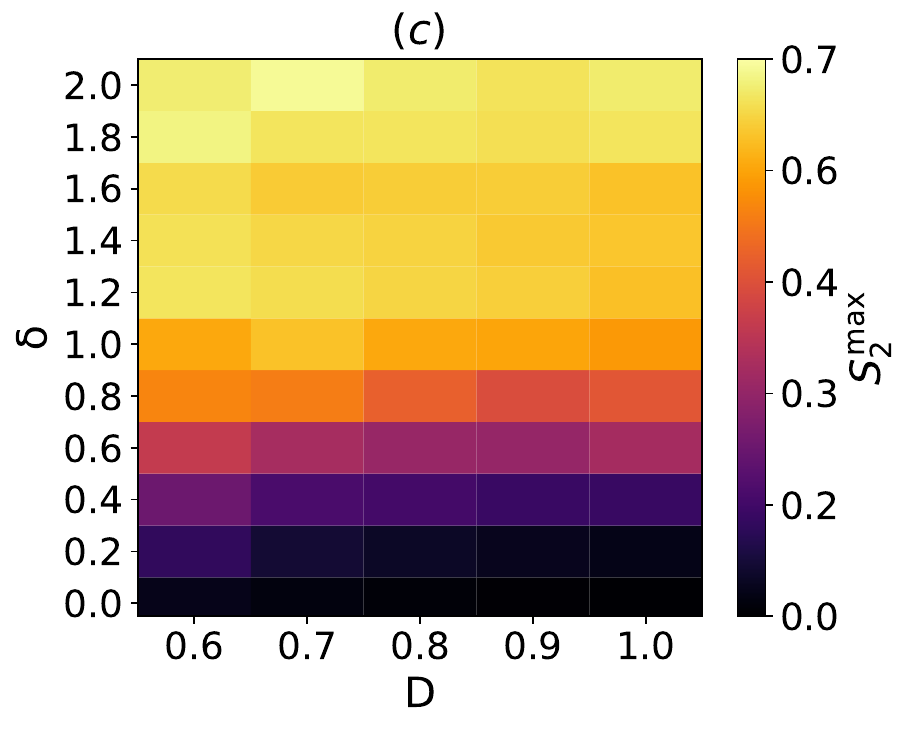}\\
  \includegraphics[width=0.68\columnwidth]{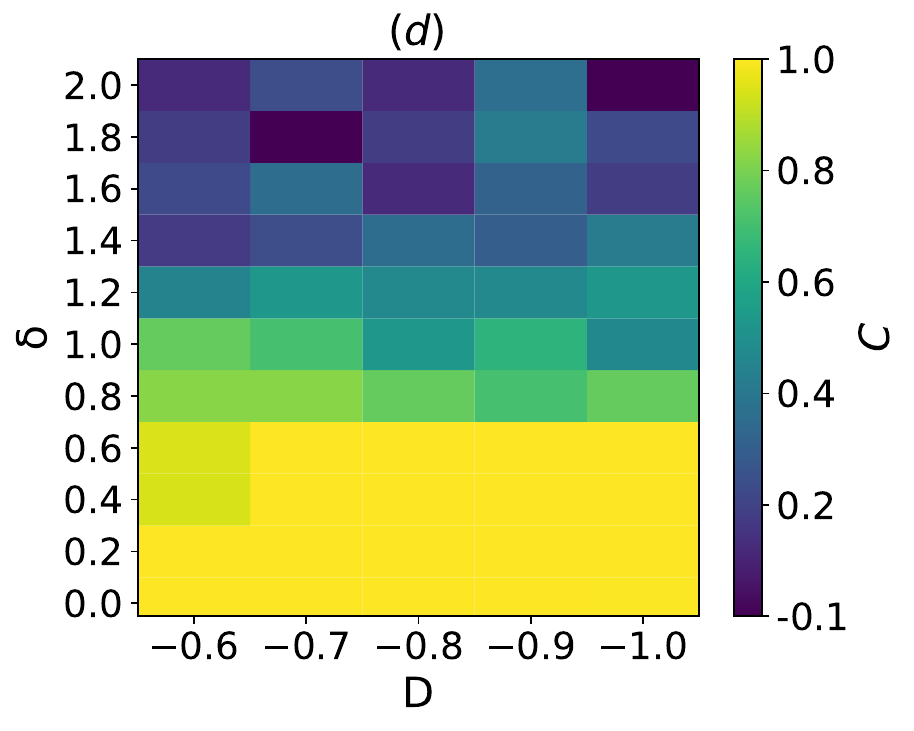}
  \includegraphics[width=0.68\columnwidth]{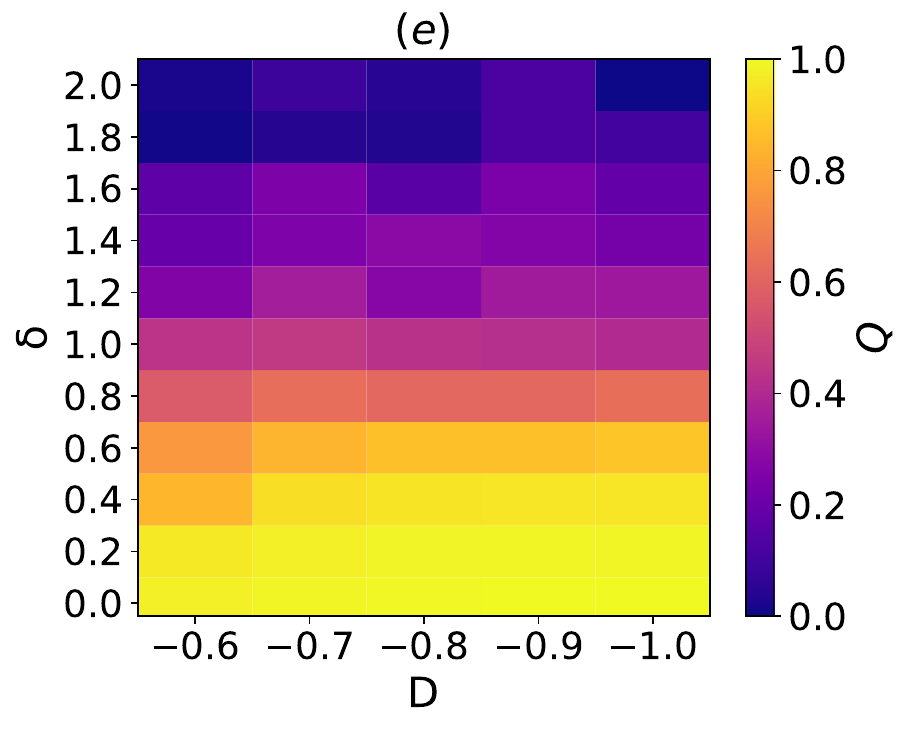}
  \includegraphics[width=0.68\columnwidth]{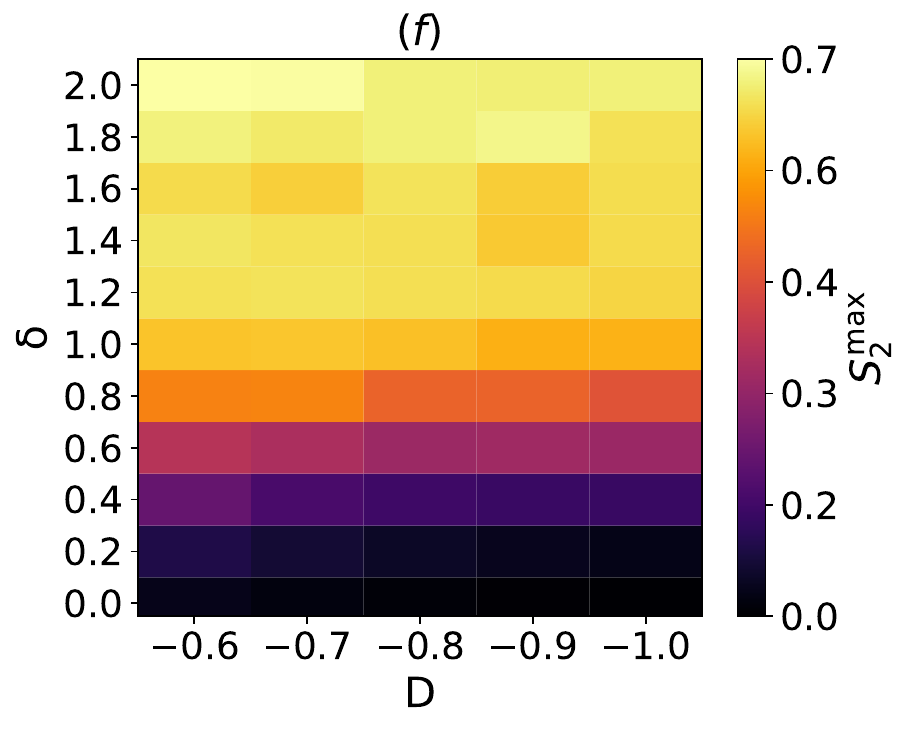}\\
  \caption{Colormaps of (a,d) skyrmion number \(C\) (b,e) skyrmion stability \(Q\) (c,f) maximum Rényi-2 entropy \(S_{2}^{\max}\) as a function of
  DMI strength $D$ and disorder strength \(\delta\) at fixed anisotropy parameter \(A = 0.3\) for system size \(L = 7\).}
  \label{fig:fixA}
\end{figure*}
While short-range entanglement reflects local quantum correlations that typically decay with distance, long-range entanglement captures the 
presence of nonlocal quantum order. In systems with strong interactions and disorder, 
long-range entanglement serves as a key indicator of possible topological phases that remain robust against local perturbations. To probe such 
global quantum correlations and long-range entanglement in our system, we evaluate the TEE. 
using the second Rényi entropy by 
partitioning the system into three regions, \( A \), \( B \), and \( C \), as illustrated in 
Fig.~\ref{fig:tee_region}. The TEE was computed using the approach proposed in 
Ref.~\cite{PhysRevLett.96.110404}, based on the combination:
\[
S_{\text{topo}} = S_{A} + S_{B} + S_{C} - S_{AB} - S_{AC} - S_{BC} + S_{ABC},
\]
where \( S_X \) denotes the Rényi-2 entropy of region \( X \).
We performed \( S_{\text{topo}}\) calculations at \( D = 0.8 \) and \( A = 0.3 \)  
for all values of disorder strength \( \delta \) and average the results over 50 disorder realizations.
\begin{figure}[H]
    \centering
    \includegraphics[width=0.94\linewidth]{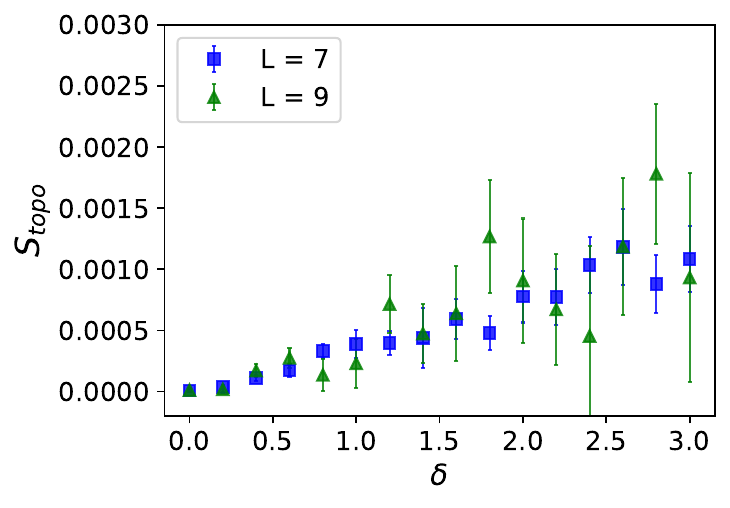}
    \caption{TEE $S_{\mathrm{topo}}$ as a function of disorder strength 
    $\delta$ calculated for lattice size \( L = 7 \) and \( L = 9 \)
    at $D = 0.8$ and $A = 0.3$ with standard error bars over 50 disorder realizations.}
    \label{fig:tee_delta}
\end{figure}

For both \( L = 7 \) and \( L = 9 \), the value of \( S_{\text{topo}} \) remains negligible for all values of \( \delta \), as shown in Fig.~\ref{fig:tee_delta}.
These results indicate that long-range entanglement is
absent in our system at all disorder strengths.
As discussed earlier (see Fig.~\ref{fig:maxRenyibatch}), the maximum Rényi-2 entropy (local quantum entanglement) decreases as the subsystem (region \( A \)) size increases at a given disorder strength.
Collectively, our calculations show that the disorder enhances local quantum entanglement while long-range entanglement remains absent across all the disorder regimes.

Negligible value of \( S_{\text{topo}} \) also shows that it is not a sensitive probe for detecting skyrmion phases in our system.
In contrast, Ref.~\cite{vijayan2023topologicalentanglemententropyidentify} reports that calculations
of TEE using the von 
Neumann entropy can detect the skyrmion phase on a triangular lattice. 
However, our findings suggest that TEE calculation using the Rényi-2 entropy on a square lattice is not sensitive in this 
context. This is consistent with prior theoretical results, which 
show that for non-chiral systems~\cite{PhysRevLett.103.261601}, the TEE is expected to be independent of the Rényi index. 
However, for chiral systems, the Rényi entropies may exhibit different behavior, and systematic 
studies in this direction remain limited, and these directions remain open for future research.

\section{Interplay of Dzyaloshinskii-Moriya Interaction, Heisenberg Anisotropy and Disorder}\label{sec:varDA}
\begin{figure*}[htb] 
  \centering
  \includegraphics[width=0.68\columnwidth]{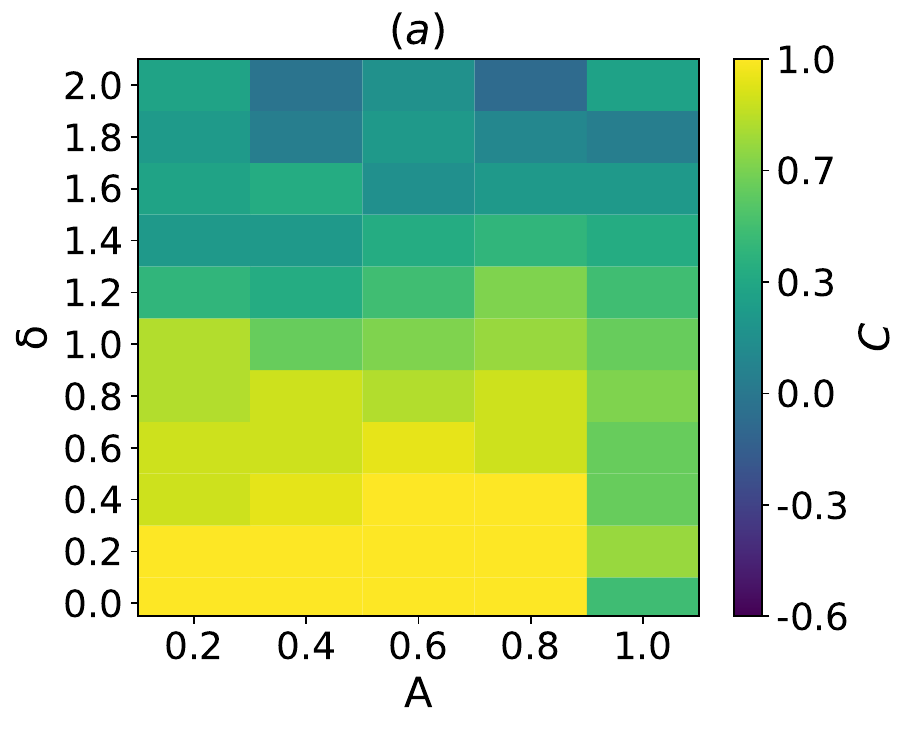}
  \includegraphics[width=0.68\columnwidth]{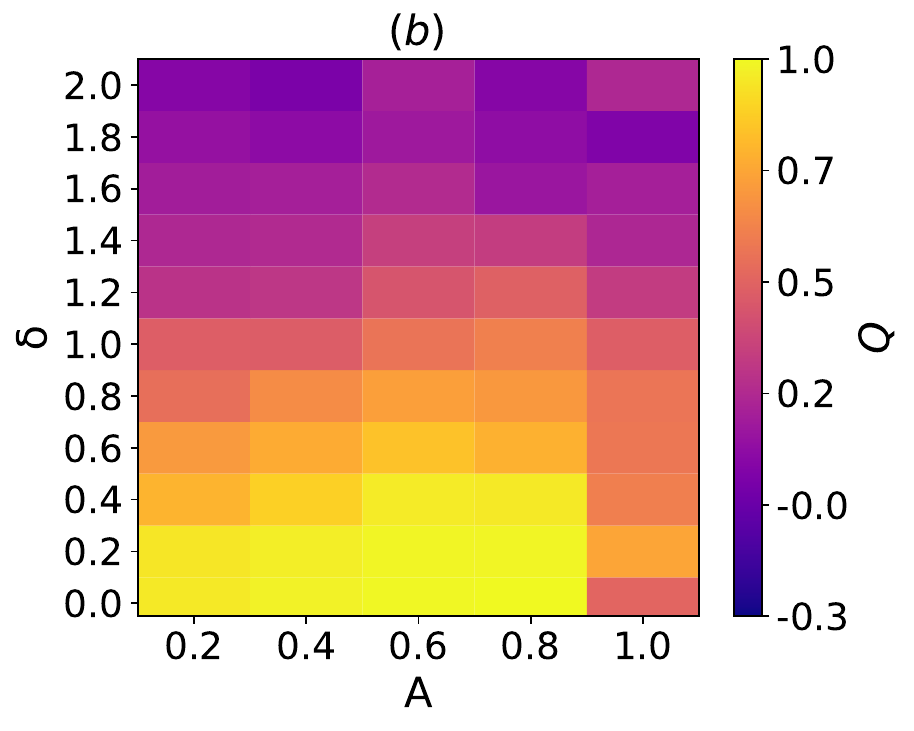}
  \includegraphics[width=0.68\columnwidth]{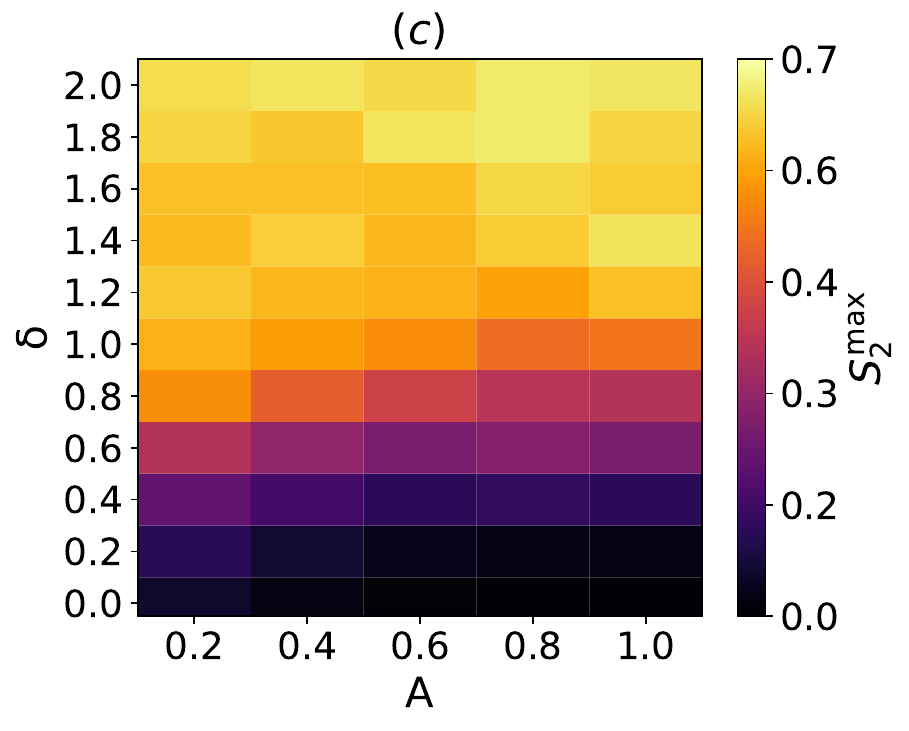}\\
  \includegraphics[width=0.68\columnwidth]{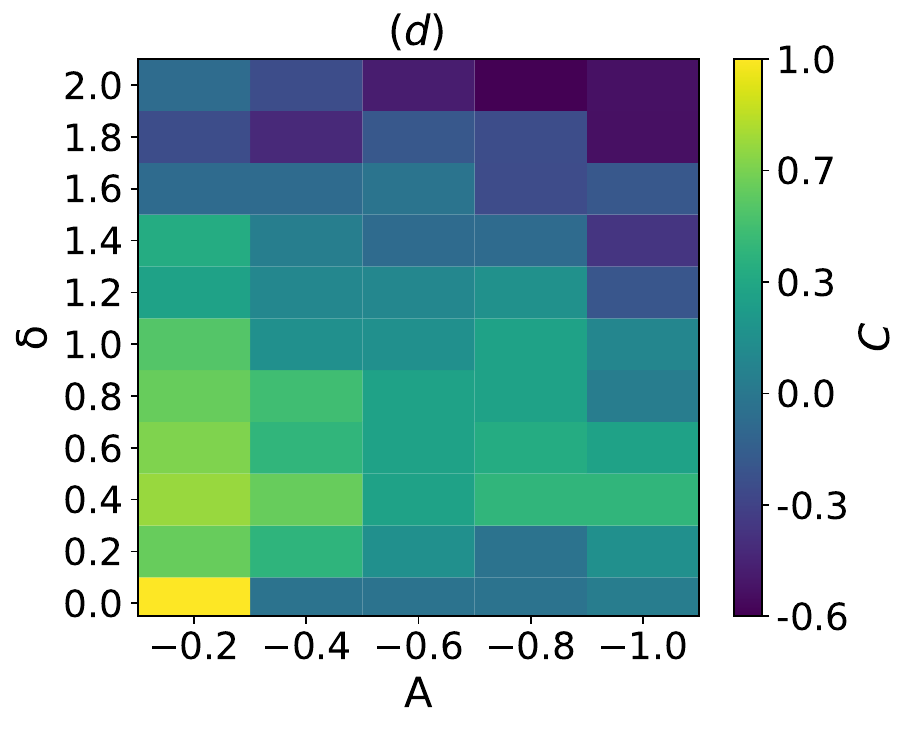}
  \includegraphics[width=0.68\columnwidth]{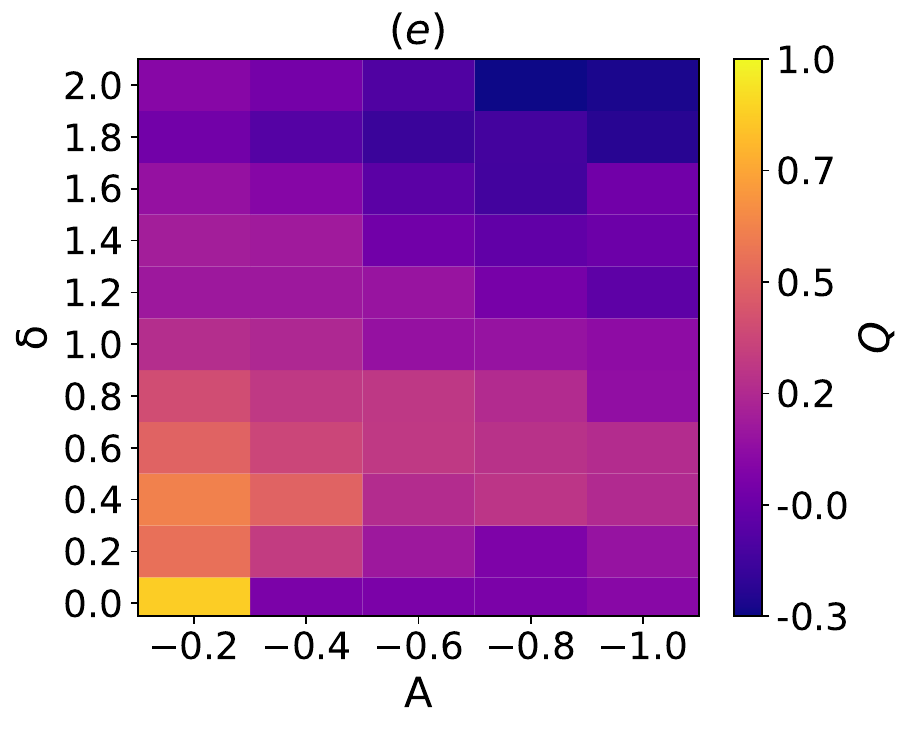}
  \includegraphics[width=0.68\columnwidth]{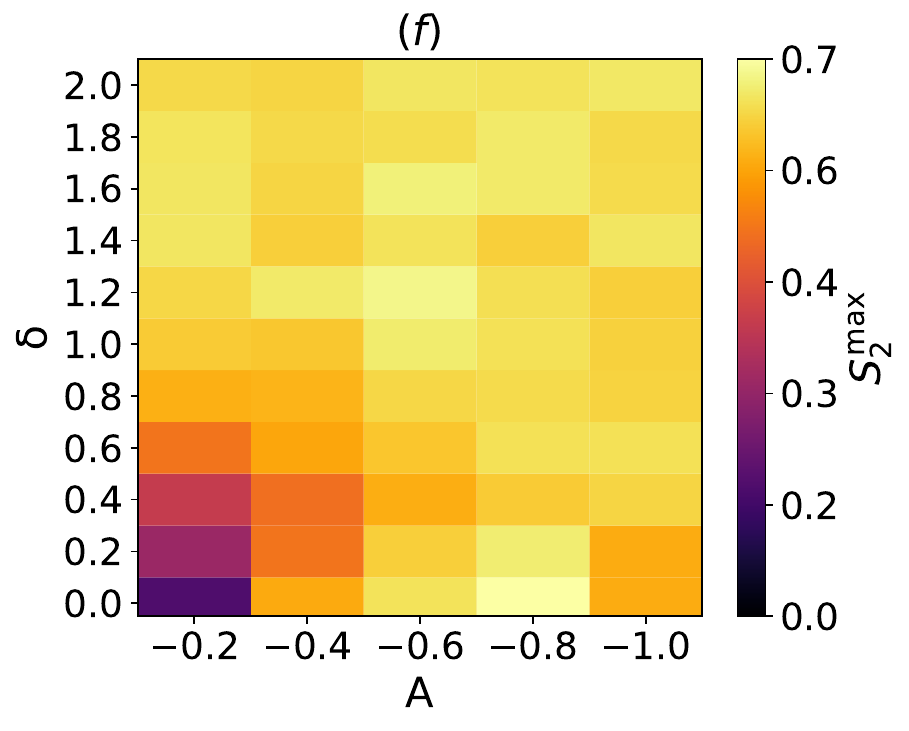}\\
  \caption{Colormaps of (a,d) skyrmion number \(C\) (b,e) skyrmion stability \(Q\) (c,f) maximum Rényi-2 entropy \(S_{2}^{\max}\) as a function of
  anisotropy parameter $A$ and disorder strength \(\delta\) at fixed DMI strength \(D = 0.7\) for system size \(L = 7\).}
  \label{fig:fixD}
\end{figure*}
The interplay between DMI, Heisenberg anisotropy, and disorder plays a crucial role in determining the stability and entanglement properties of quantum skyrmions. 
While DMI favors chiral spin textures and Heisenberg anisotropy helps stabilize the skyrmion spin texture, disorder introduces competing randomness that can distort or even degrade these textures. 
Understanding how these factors collectively influence skyrmion formation, their stability, and entanglement properties is essential for identifying regimes where skyrmions remain robust against 
these random bond disorders. 
In the following, we first explore this competition by analyzing how varying DMI strength \(D\) in the presence of disorder affects the stability of skyrmions and their entanglement properties.
To investigate this,
we fix the anisotropy parameter to \(A=0.3\) and the lattice size to \(L=7\). 
Within this setup, we systematically vary the disorder strength \(\delta\) in the range \(0.0 \leq \delta \leq 2.0\) with increments of \(0.2\). 
In contrast, the range of variation of the DMI strength is chosen such that we obtain a single stable quantum skyrmion at \(\delta = 0\). 
Thus, we varied \(D\) from 0.6 to 1.0 in increments of 0.1, and this range is large enough to study the effect of varying DMI on these properties. Finally, we average all the observables over 15 disorder realizations.

To examine the effect of DMI strength on the skyrmion number \(C\), we plot \(C\) as a function of DMI strength and disorder strength in Fig.~\ref{fig:fixA}(a). 
At zero disorder, the skyrmion number remains unity for all values of \(D\). 
As the disorder strength increases, \(C\) deviate from unity 
at lower values of $\delta$ for smaller values of \(D\), while larger \(D\) values maintain unity skyrmion number over a wider range of disorder strengths. 

Now, to study the stability of these quantum skyrmions, we plot the skyrmion stability \(Q\) as a function of disorder strength and DMI strength in Fig.~\ref{fig:fixA}(b). 
At smaller values of \(D\), stability decreases at weaker disorder strength, while at larger values of \(D\), skyrmions remain stable up to stronger disorder strengths. 
This behavior can be understood from the fact that, at small \(D\), the skyrmion state is comparatively less favorable \cite{Mohylna_2021}, 
which makes it more fragile against disorder. By contrast, larger values of \(D\) with enhanced skyrmion stability allow the system to sustain stable skyrmions up to higher disorder.  

We further analyze variation of entanglement in Fig.~\ref{fig:fixA}(c), where the maximum Rényi-2 entropy, \(S_{2}^{\max}\), is shown as a function of disorder strength and DMI strength. 
For small \(D\), the \(S_{2}^{\max}\) rises rapidly once disorder is introduced, while for larger \(D\) the increase occurs only at higher disorder strengths. 
This behavior shows that, at low \(D\), the system develops quantum entanglement more quickly in response to disorder, following the trends similar to that of $C$, which degrades earlier at lower values of \(D\). 
As discussed in Sec.~\ref{sec:entanglement}, a similar trend is observed for the spin magnitude range, \(\Delta S\) (not shown here). 
At small \(D\), \(\Delta S\) increases more rapidly with disorder compared to larger \(D\). 
This reflects that larger deviations from the average spin magnitude emerge at weaker disorder when \(D\) is small, 
which again reduces skyrmion stability in this regime.

Changing the sign of \(D\) reverses the chirality of the DMI, which in turn determines the rotational sense of the spin texture ~\cite{CAMLEY2023100605}. 
Although the magnitude of \(D\) controls the skyrmion stability, its sign dictates the handedness of the skyrmion configuration. 
Studying negative \(D\) thus allows us to verify that physical quantities, such as stability and entanglement, remain unaffected under reversal of chirality.
Thus, we extend our analysis to negative values of \(D\). 
In Fig.~\ref{fig:fixA}(d), we show the skyrmion number \(C\) for \(D\) ranging from \(-0.6\) to \(-1.0\) in steps of \(-0.1\).
We observe that \(C\) is unity for all values of \(D\) at \(\delta = 0.0\) and starts deviating from unity at higher disorder strengths
for larger negative values of \(D\). 
Similarly, the skyrmion stability shown in Fig.~\ref{fig:fixA}(e) 
decreases at higher values of $\delta$ for larger negative values of \(D\). 
The corresponding entanglement shown in Fig.~\ref{fig:fixA}(f) exhibits the same qualitative behavior as \(C\), 
and its growth is suppressed and shifted to higher values of disorder strength \(\delta\) for larger negative values of \(D\). 
In summary, we find that the behavior of \(C\), \(Q\), and \(S_{2}^{\max}\) is more or less the same for both positive and negative values of \(D\), confirming that 
these properties remain unaffected under the reversal of chirality.


Now we explore the competition by analyzing how varying the anisotropy parameter \(A\) and disorder strength \(\delta\) affect the skyrmion stability and entanglement properties.
For this,
we fix the DMI strength to \(D=0.7\) and the lattice size to \(L=7\).
We then systematically vary the disorder strength \(\delta\) in the range \(0.0 \leq \delta \leq 2.0\) with increments of \(0.2\). 
At the same time, the range of variation of the anisotropy parameter is chosen such that we obtain a single stable quantum skyrmion at \(\delta = 0\), and the range is large enough to study the effect of the anisotropy parameter on these properties. 
Thus, we varied \(A\) from 0.2 to 1.0 in increments of 0.2, and finally averaged all the observables over 15 disorder realizations.

In Fig.~\ref{fig:fixD}(a), we show the skyrmion number \(C\) as a function of disorder strength and anisotropy parameter. 
At zero disorder, the skyrmion number remains unity for all values of \(A\). 
As disorder increases, \(C\) deviates from the unity value, first at smaller values of \(A\), while larger \(A\) maintains stability over a wider range of disorder. 
However, for the largest anisotropy value, \(A = 1.0\), we observe that \(C\) is not unity at $\delta = 0$, suggesting the absence of a skyrmion.

Now, to examine the stability of these skyrmions, we present the variation of skyrmion stability \(Q\) as a function of \(\delta\) and \(A\) in Fig.~\ref{fig:fixD}(b). 
We find that the stability decreases once disorder is introduced. At small values of \(A\), this reduction occurs rapidly, 
while for larger \(A\) skyrmions remain stable up to stronger disorder. This behavior arises because the skyrmion state is 
less favorable for small anisotropy, making it more fragile against disorder. 
At a larger anisotropy, stability improves; however, at \(A = 1.0\), stability is very small, indicating the absence of skyrmion. 
To better understand this behavior, we next analyze how entanglement evolves at these values of \(A\).  
The maximum Rényi-2 entropy \(S_{2}^{\max}\), shown in Fig.~\ref{fig:fixD}(c), provides insight into the entanglement. 
At small values of \(A\), \(S_{2}^{\max}\) increases rapidly with disorder, whereas for larger anisotropy the increase sets in only at stronger disorder,
following the trends similar to that of \(C\). 
Interestingly, at \(A=1.0\), the maximum value of \(S_{2}^{\max}\) remains very low, signaling a highly ordered phase. 
Inspection of the spin configurations confirms this to be a ferromagnetic state, consistent with the non-unity skyrmion number and small value of skyrmion stability observed in Fig.~\ref{fig:fixD}(a) and (b), 
and the fact that a larger anisotropy favors more out-of-plane spin alignment, which results in a ferromagnetic phase.  
Similarly, as discussed earlier in Sec.~\ref{sec:entanglement}, spin magnitude range \(\Delta S\) (not shown here) also follows the same trend as entanglement. 
For small anisotropy, \(\Delta S\) increases more rapidly with disorder compared to larger \(A\). 
This behavior indicates that larger deviations from the average spin magnitude appear earlier at weak anisotropy, 
which again contributes to reduced skyrmion stability. 
Notably, at \(A=1.0\), \(\Delta S\) follows a gradual increase with disorder similar to \(S_{2}^{\max}\). 
Suggesting that the ferromagnetic phase has similar stability against disorder, as skyrmions at large anisotropy ($A = 0.8$).  

The sign of the anisotropy parameter plays a crucial role in stabilizing skyrmions by favoring spin alignment along a preferred direction. 
While positive anisotropy favors skyrmion formation, negative anisotropy suppresses skyrmion formation. 
Studying negative anisotropy, therefore, allows us to investigate its influence on skyrmion formation, stability, and entanglement characteristics.
Thus, we repeated our analysis for negative values of \(A\).
In Fig.~\ref{fig:fixD}(d--f), we present colormaps of skyrmion number, skyrmion stability, and maximum Rényi-2 entropy, 
for anisotropy values ranging from \(-0.2\) to \(-1.0\) in steps of \(-0.2\). 
We observe non-unity skyrmion number and stability (nearly zero) for most negative values of \(A\), except at \(A=-0.2\), where a skyrmion appears in the ground state. 
For \(A\) between \(-0.8\) and \(-0.4\), spin configurations at \(\delta=0\) reveal in-plane alignment without any specific order, which also corresponds to a high entanglement state. 
At \(A=-1.0\), the entanglement is slightly lower compared to the intermediate values, and inspection of the spin textures confirms that the system realizes a helical phase which is slightly more ordered than in-plane alignment.
In summary, we find that changing the sign of the anisotropy parameter highly suppresses skyrmion formation and makes them more vulnerable to disorder in contrast to positive anisotropy.  

\section{Conclusion}
In this work, we studied the properties of quantum
skyrmions under random bond disorder
using a neural network quantum state approach. We observe that disorder reduces the stability of quantum skyrmions, 
ultimately causing them to collapse at high disorder strengths. 
We used the second Rényi entropy to study the
entanglement, interestingly, we found that increasing disorder
leads to enhanced local quantum entanglement and ultimately leads to its saturation
at higher disorder strength. 
We also found that the topological entanglement entropy, calculated using the second Rényi entropy, 
remains negligible at all 
disorder strengths, indicating, first, the absence of long-range entanglement at all disorder strengths
and second, that it is 
not a reliable probe to characterize skyrmion phases.
These findings contribute to a deeper understanding of
disorder-induced behavior of quantum skyrmions and
provide guidance for using disorder as a constructive tool for quantum materials.

\begin{acknowledgments}
We acknowledge National Supercomputing Mission (NSM) for providing computing resources of 
‘PARAM RUDRA’ at S.N. Bose National Centre for Basic Sciences, which is implemented by 
C-DAC and supported by the Ministry of Electronics and Information Technology (MeitY) and 
Department of Science and Technology (DST), Government of India, and the use of CCS3 
computing cluster at SINP.
\end{acknowledgments}

\appendix
\section{NQS Implementation and Training Details}
\begin{figure}[htb]
    \centering
    \includegraphics[width=0.45\linewidth]{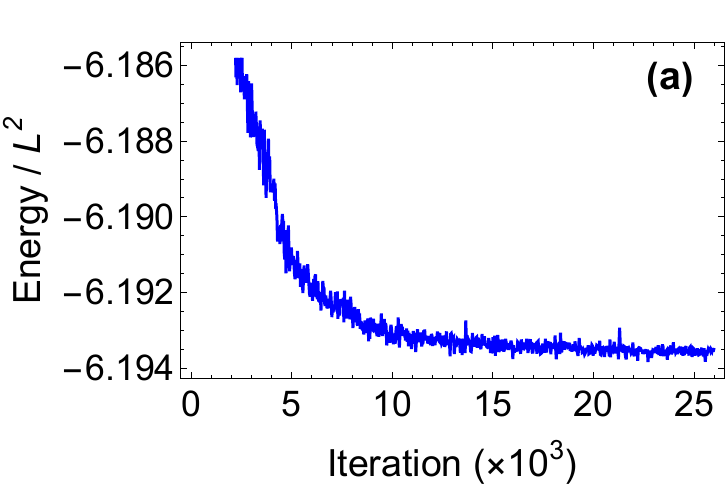}
    \includegraphics[width=0.45\linewidth]{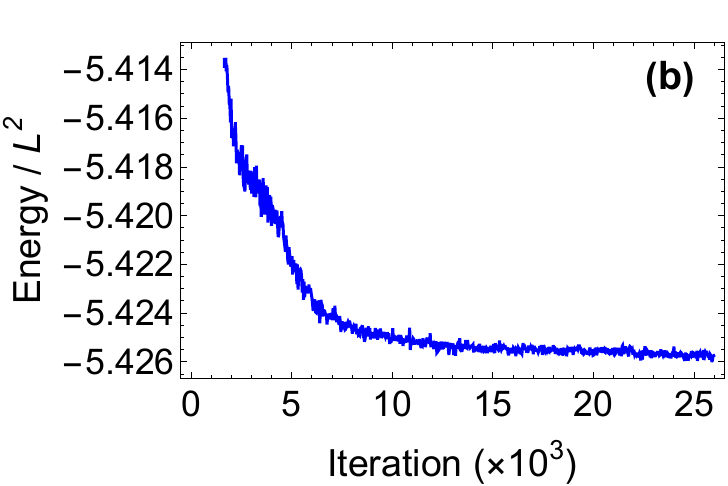}
    \caption{Energy convergence obtained using NQS shows the total energy per lattice site with 
    standard errors for $\delta = 0.0$, 
    $D = 0.8$, and $A = 0.3$. For lattice size (a)$L = 7$ and (b)$L = 9$.}
    \label{fig:energy_convergence_small}
\end{figure}
In this appendix, we provide the technical details of our NQS implementation. 
Specifically, we describe how the expectation values of observables are computed and the procedures used for training the network. 
Given a variational wave function $|\psi_\theta\rangle$, the expectation value of an 
operator $\hat{O}$ can be written as:
\begin{equation}
\langle \hat{O} \rangle 
= \sum_{\sigma} p_\theta(\sigma)\, O^{\text{loc}}_\theta(\sigma),
\label{eq:exp_value}
\end{equation}
where the probability distribution associated with configurations is given by:
\begin{equation}
p_\theta(\sigma) = \frac{|\psi_\theta(\sigma)|^2}{\sum_{\sigma} |\psi_\theta(\sigma)|^2},
\label{eq:prob_dist}
\end{equation}
with $\psi_\theta(\sigma) = \langle \sigma|\psi_\theta\rangle$
and the corresponding local estimator of the operator is
\begin{equation}
O^{\text{loc}}_\theta(\sigma) 
= \sum_{\sigma'} \frac{\langle \sigma|\hat{O}|\sigma'\rangle \, \psi_\theta(\sigma')}{\psi_\theta(\sigma)}.
\label{eq:local_est}
\end{equation}
Since summing 
over all configurations \( |\sigma\rangle \) becomes infeasible for large systems, we employ Markov 
Chain Monte Carlo sampling (using the Metropolis-Hastings algorithm) to approximate the expectation as
\begin{equation}
\langle \hat{O} \rangle \approx \frac{1}{N} \sum_{n=1}^{N} O^{\text{loc}}_{\theta}(\sigma_n),
\end{equation}
where \( \{ \sigma_n \} \) are sampled configurations and \( N \) is the total number of samples.
During training, we used \(10^4\) Monte Carlo samples for estimating the 
energy, while \(10^7\) samples were used for computing other observables~\cite{PhysRevB.108.094410}.

For optimization, we used the Adam optimizer with momentum parameters \( \beta_{1} = 0.9 \) and \( \beta_{2} = 0.999 \) \cite{kingma2017adammethodstochasticoptimization}. 
Separate learning rate schedules were used for different parameter blocks. For the 
\( \rho \)-parameters, the initial schedule consisted of a linear warm-up from 0 to 0.001 over the 
first 5000 steps, followed by a constant rate of 0.001 until step 22000, and then a reduced 
rate of 0.0001 thereafter. For the \( \phi \)-parameters, the learning rate was kept at 0.001 
until step 22000 and subsequently decreased to 0.0001.
Simulations were run for 26000 iterations to examine the energy convergence behavior for all system sizes. 
Figure~\ref{fig:energy_convergence_small} presents the convergence curves for \( \delta = 0.0 \) at \( D = 0.8 \) and \( A = 0.3 \) for the lattice size \( L = 7 \) and \( L = 9 \).
After analyzing the convergence trends, the schedules were adjusted: for \( L = 7 \), the warm-up was extended from 5000 to 5800 steps, the second transition point 
was reduced from 22000 to 16000 steps, and training proceeded for 19000 iterations; for \( L = 9 \), 
the warm-up was extended to 7000 steps, the second transition point was reduced to 17500 steps, and 
training ran for 20000 iterations.

\label{app:NQS}

\bibliographystyle{apsrev4-2}
\bibliography{references}

\end{document}